\documentclass[preprint,prd,nofootinbib,tightenlines,amssymb]{revtex4}
\usepackage[dvipdfmx]{graphicx}
\usepackage{bm}
\usepackage{amsmath}
\usepackage{hhline}
\usepackage{color}
\usepackage{xcolor}
\usepackage{slashed}
\usepackage{cancel}
\usepackage{relsize}
\usepackage[T1]{fontenc}
\usepackage{upquote}
\usepackage{changes} 

\usepackage{ftnxtra}

\newcommand{\Eq}{&\hs{-0.2cm}=\hs{-0.2cm}&}

\newcommand{\white}[1]{{\color[rgb]{1,1,1} #1}}
\newcommand{\black}[1]{{\color[rgb]{0,0,0} #1}}
\newcommand{\bs}[1]{{\boldsymbol{#1}}}
\newcommand{\hs}[1]{{\hspace{#1}}}
\newcommand{\vs}[1]{{\vspace{#1}}}
\newcommand{\tf}[1]{{\textsf{#1}^{}}}
\newcommand{\tx}[1]{{\text{#1}^{}}}
\newcommand{\nn}{\nonumber\\}

\newcommand{\sx}[2]{{\scalebox{#1}{#2}}}
\newcommand{\dd}{{\mathrm{d}}}

\newcommand{\XXXNbNb}
{X\hs{-0.03cm}X\hs{-0.03cm}X \to \bar{N}\hs{-0.03cm}\bar{N}}
\newcommand{\XXNXbNb}
{X\hs{-0.03cm}X\hs{-0.03cm}N \to \bar{X}\hs{-0.03cm}\bar{N}}
\newcommand{\XNNXbXb}
{X\hs{-0.03cm}N\hs{-0.03cm}N \to \bar{X}\hs{-0.03cm}\bar{X}}

\newcommand{\XbXbXbNN}
{\bar{X}\hs{-0.03cm}\bar{X}\hs{-0.03cm}\bar{X} \to N\hs{-0.03cm}N}
\newcommand{\XbXbNbXN}
{\bar{X}\hs{-0.03cm}\bar{X}\hs{-0.03cm}\bar{N} \to X\hs{-0.03cm}N}
\newcommand{\XbNbNbXX}
{\bar{X}\hs{-0.03cm}\bar{N}\hs{-0.03cm}\bar{N} \to X\hs{-0.03cm}X}

\newcommand{\XXXXbXb}
{X\hs{-0.03cm}X\hs{-0.03cm}X \leftrightarrow \bar{X}\hs{-0.03cm}\bar{X}}
\newcommand{\XbXbXbXX}
{\bar{X}\hs{-0.03cm}\bar{X}\hs{-0.03cm}\bar{X} \leftrightarrow X\hs{-0.03cm}X}

\newcommand{\NNbXXb}
{N\hs{-0.03cm}\bar{N} \to X\hs{-0.03cm}\bar{X}}
\newcommand{\XXbNNb}
{X\hs{-0.03cm}\bar{X} \to N\hs{-0.03cm}\bar{N}}

\newcommand{\XX}{X\hs{-0.03cm}\bar{X}}
\newcommand{\NN}{N\hs{-0.03cm}\bar{N}}

\newcommand{\NNXXX}{\bar{N}\hs{-0.03cm}\bar{N} \to X\hs{-0.03cm}X\hs{-0.03cm}X}
\newcommand{\XXNNNN}{X\hs{-0.03cm}\bar{X} \to N\hs{-0.03cm}\bar{N}\hs{-0.03cm}N\hs{-0.03cm}\bar{N}}

\newcommand{\lambdaXS}{\lambda_{X\hs{-0.03cm}S}}
\newcommand{\Zf}{\mathbb{Z}^{}_4}

\newcommand{\YN}{Y_{\white{\bar{\black{N}}}}}
\newcommand{\YX}{Y_{\white{\bar{\black{X}}}}}
\newcommand{\YNb}{Y_{\bar{N}}}
\newcommand{\YXb}{Y_{\bar{X}}}

\newcommand{\YeqN}{Y^0_{\white{\bar{\black{N}}}}}
\newcommand{\YeqNb}{Y^0_{\bar{N}}}
\newcommand{\YeqX}{Y^0_{\white{\bar{\black{X}}}}}
\newcommand{\YeqXb}{Y^0_{\bar{X}}}

\newcommand{\etaDM}{\eta^{}_{\white{\bar{\tf{\black{DM}}}}}}
\newcommand{\etaB}{\eta^{}_{\white{\bar{\tf{\black{B}}}}}}
\newcommand{\etaN}{\eta^{}_{\white{\bar{\black{N}}}}}
\newcommand{\etaX}{\eta^{}_{\white{\bar{\black{X}}}}}
\newcommand{\epsN}{\epsilon^{}_{\white{\bar{\black{N}}}}}

\newcommand{\muDM}{\mu_{\white{\overline{\black{\tf{DM}}}}}}
\newcommand{\muDMb}{\mu_{\overline{\tf{DM}}}}
\newcommand{\muN}{\mu_{\white{\bar{\black{N}}}}}
\newcommand{\muNb}{\mu_{\bar{N}}}
\newcommand{\muX}{\mu_{\white{\bar{\black{X}}}}}
\newcommand{\muXb}{\mu_{\bar{X}}}

\newcommand{\muZ}{\mu_{\white{\bar{\black{Z}}}}}
\newcommand{\muZb}{\mu_{\bar{Z}}}

\newcommand{\XXXZbZb}
{X\hs{-0.03cm}X\hs{-0.03cm}X \leftrightarrow \bar{Z}\hs{-0.02cm}\bar{Z}}
\newcommand{\XXZXbZb}
{X\hs{-0.03cm}X\hs{-0.03cm}Z \leftrightarrow \bar{X}\hs{-0.05cm}\bar{Z}}
\newcommand{\XZZXbXb}
{X\hs{-0.05cm}Z\hs{-0.02cm}Z \leftrightarrow \bar{X}\hs{-0.03cm}\bar{X}}

\begin{document}
\baselineskip=14.5pt \parskip=2.5pt

\vspace*{3em}

\preprint{KIAS-P22058}

\title{\large An Asymmetric SIMP Dark Matter Model}

\author{Shu$\,$-Yu\,\,Ho\footnote[1]{phyhunter@kias.re.kr}}

\affiliation{
Korea Institute for Advanced Study, Seoul 02455, Republic of Korea
\vspace{3ex}}

\begin{abstract}
In this paper, we construct the first asymmetric strongly interacting massive particles (SIMP) dark matter (DM) model, where a new vector-like fermion and a new complex scalar both having nonzero chemical potentials can be asymmetric DM particles.\,\,After the spontaneous breaking of a U$(1)^{}_\tf{D}$ dark gauge symmetry, these two particles can have accidental $\,\Zf$ charges making them stable.\,\,By adding one more complex scalar as a mediator between the SIMP DM, the relic density of DM is determined by $3 \to 2$ and two-loop induced $2 \to 2$ annihilations in this model.\,\,On the other hand, the SIMP DM can maintain kinetic equilibrium with the thermal bath until the DM freeze-out temperature via the new gauge interaction.\,\,Interestingly, this model can have a bouncing effect on DM, whereby the DM number density rises after the chemical freeze-out of DM.\,\,With this effect, the prediction of the DM self-interacting cross section in this model can be consistent with astrophysical observations, and the ratio of the DM energy density to the baryonic matter energy density can be explained by primordial asymmetries.\,\,We also predict the DM-electron elastic scattering cross section that can be used to test this model in future projected experiments.
\end{abstract}

\maketitle

\section{Introduction}\label{sec:1}

The existences of dark matter (DM) and residual ordinary matter (baryon asymmetry) in the present universe are two fascinating problems in modern cosmology since they cannot be accommodated well with the standard model (SM) of particle physics.\,\,The former contributes about 25\% to the  energy density of the present universe, and the latter only constitutes 5\% or so.\,\,Without the anthropic principle, it may be simply a cosmological accidence that these two distinct matter densities are comparable.\,\,Nonetheless, it is hard for physicists not to speculate that they come from a similar origin as the densities of DM and baryons are just different by a factor of about five, which provides another hint for physics beyond the SM.

To this end, a lot of attempts have been proposed to address these problems.\,\,For instance, weakly interacting massive particles (WIMP)~\cite{Lee:1977ua} is one of the promising DM candidates people have drawn attention to over the last decade.\,\,In the WIMP scenario, the chemical potentials of DM and anti-DM, $\muDM$ and $\muDMb$, are assumed to be zero,\footnote{If DM particles are in chemical equilibrium with the thermal bath, for the annihilation process $\tf{DM} + \overline{\tf{DM}} \leftrightarrow e^+ + e^-$, we have $\muDM + \muDMb = \mu_{e^+} + \mu_{e^-}$.\,\,On the other hand, the inelastic scattering $e^-e^- \leftrightarrow e^-e^-\gamma$ and pair annihilation $e^+e^- \leftrightarrow 2\gamma$ have reaction rates much bigger than the expansion rate of the universe.\,\,Thus, we can obtain the solution $\mu_{e^+} + \mu_{e^-}  = \mu_\gamma = 0$, from which $\muDM + \muDMb = 0$.} then the relic density of DM is set by the annihilation rate of a DM-anti-DM pair into SM particles.\,\,Since the typical mass range of WIMP DM is from $\sim$\,1\,GeV to $\sim$\,100\,TeV, hence we can detect WIMP DM directly through WIMP-nucleon interactions.\,\,On the other hand, various feasible scenarios can account for the baryon asymmetry such as Affleck-Dine baryogenesis~\cite{Affleck:1984fy}, baryogenesis via leptogenesis~\cite{Fukugita:1986hr}, and electroweak baryogenesis~\cite{Kuzmin:1985mm}.\,\,However, the above scenarios consider different underlying origins to deal with the DM and the baryon asymmetry independently, which still leaves the observed DM-to-baryon density ratio, $\Omega^{}_\tf{DM}/\Omega^{}_\tf{B} \simeq 5$, as a cosmological coincidence problem.

Asymmetric DM (ADM) is an alternative scenario of DM which can explain the coincidence of the present DM-to-baryon density ratio~\cite{Kaplan:2009ag,Iminniyaz:2011yp,Graesser:2011wi,Ghosh:2020lma}.\,\,In comparison with the WIMP paradigm, the chemical potentials of DM and anti-DM in the ADM scenario are not zero, but $\muDM = -{}^{}{}^{}\muDMb \neq 0$.\,\,As a result, the DM relic abundance is determined by a conserved DM asymmetry quantity rather than the annihilation cross section of DM if the DM is fully asymmetric.\,\,In other words, the DM asymmetry $\etaDM\hs{-0.05cm}$ is produced by the same physical mechanism as the baryon asymmetry $\etaB$.\,Therefore, we can obtain $\Omega^{}_\tf{DM}/\Omega^{}_\tf{B} \simeq (m^{}_\tf{DM}/m_p)/(\etaDM/\etaB) \simeq 5$ if $m^{}_\tf{DM} \simeq 5{}^{}m_p \simeq 5\,\tx{GeV}\hs{-0.05cm}$ and $\etaDM \simeq \etaB\hs{-0.05cm}$ with $m^{}_\tf{DM}{}^{}(m_p)$ being the mass of DM\,(proton).\,\,However, the current direct detection experiments have narrowed the testable region for the DM-nucleon cross sections with the DM masses above $\sim$\,5\,GeV~\cite{XENON:2019gfn,Workman:2022ynf,LUX-ZEPLIN:2022qhg}.\,\,Thus, it may be challenging to examine the ADM scenario by the direct detection searches if we receive no DM-nucleon scattering events in the future unless the ADM mass (DM asymmetry) is much lighter than 5\,GeV (higher than $\etaB$). 

Strongly interacting massive particles (SIMP) is an appealing DM scenario which has gained attention since it predicts strong couplings and low masses of DM that can resolve some small-scale issues in astrophysics such as the core-vs-cusp problem and too-big-to-fail problem~\cite{Hochberg:2014dra}.\,\,In the SIMP scenario, the DM relic density is set by the $3 \to 2$ self-interaction of the DM, which leads to the DM with an order of unity coupling and tens of MeV to sub-GeV mass.\,\,In addition, the SIMP DM must have interactions with the SM particles to put a stop to the heat up of the DM due to the $3 \to 2$ processes, which means that we can also detect the SIMP DM by direct searches like the WIMP DM.\,\,Now, with these enticing features mentioned above, the question we may ask is can we make the SIMP DM asymmetric${}^{}{}^{}$?\,\,To answer this question, let us take a five-point interaction of the DM, ${\cal O}_X = X^5$, as an example, where $X$ is a complex scalar DM particle with a discrete symmetry.\,\,Given this interaction, the $3 \leftrightarrow 2$ processes we can have are $\XXXXbXb$ and $\XbXbXbXX$.\,\,Thus, the equations for the chemical potentials of DM and anti-DM during the chemical equilibrium epoch are $3{}^{}\muX = 2{}^{}\muXb$ and $3{}^{}\muXb = 2{}^{}\muX$ which have a unique solution $\muX = \muXb = 0$ making the SIMP DM symmetric.

In order to have an asymmetric SIMP ($a$SIMP) scenario, we observe that at least two SIMP DM particles are needed.\,\,To demonstrate this point, we consider another five-point interaction of the DM, ${\cal O}_{X\hs{-0.03cm}Z} = X^3\hs{-0.03cm}Z^2$, where $Z$ can be a complex scalar or a fermion.\,${}^{}$With this interaction, all the possible $3 \leftrightarrow 2$ processes are $\XXXZbZb{}^{}$, $\XXZXbZb{}^{}$, and $\XZZXbXb$\,(as well as the conjugate processes).\,\,Hence, during the chemical equilibrium period, we have $3{}^{}\muX = 2{}^{}\muZb$, $2{}^{}\muX + \muZ = \muXb + \muZb$, $\muX + 2{}^{}\muZ = 2{}^{}\muXb$, and their conjugate equations.\,\,These equations can be reduced to $\muX = -{}^{}\muXb$, $\muZ = -{}^{}\muZb{}^{}$, $3{}^{}\muX = 2{}^{}\muZb{}^{}$, and $3{}^{}\muXb = 2{}^{}\muZ$ which allow the DM particles to have nonzero chemical potentials such that the SIMP DM can be asymmetric.\,\,For the case where the SIMP DM particles have zero chemical potentials, $\mu_{X,{}^{}\bar{X},{}^{}Z,{}^{}\bar{Z}} = 0$, see Refs.\,\cite{Ho:2021ojb,Ho:2022erb}.\footnote{For the multi-component SIMP DM scenario, one can also refer to \cite{Choi:2021yps}.}

With this finding, the aim of this paper is concrete and straightforward.\,${}^{}$We want to build a UV complete model to realize the $a$SIMP scenario and to see its phenomenology.\,\,In particular,
this DM scenario can have a bouncing effect on DM due to the $3 \to 2$ annihilations of different species of SIMP DM.\,${}^{}$With this effect, the number density of one of the SIMP particles can rise after the DM freeze-out temperature and becomes the dominant DM component.\,\,Then, if this SIMP particle has a large DM asymmetry, we can interpret the DM-to-baryon ratio.\,\,Moreover,
we notice that the $3 \to 2$ annihilations in the $a$SIMP scenario can always generate the $2 \to 2$ annihilations at the two-loop level, affecting the thermal history of the SIMP DM.\,\,The details of these effects will be discussed in the later sections.

The outline of this paper is as follows.\,\,In the next section, we consider a pre-built model to realize the $a$SIMP scenario and briefly write down the relevant interactions and masses for the new particles.\,${}^{}$In Sec.\,\ref{sec:3}, we show the formulas for the annihilation cross sections of the $3 \to 2$ and $2 \to 2$ processes in this model.\,\,In Sec.\,\ref{sec:4}, we take into account all possible theoretical and observational constraints on this model.\,\,In Sec.\,\ref{sec:5}, we compute the DM relic density and discuss the bouncing effect of DM.\,\,In Sec.\,\ref{sec:6}, we show our predictions of the DM self-interacting cross section and the DM-electron elastic scattering cross section in this model.\,\,The last section is devoted to discussion and conclusions.

\section{$\bs{a}$SIMP model}\label{sec:2}

To achieve the $a$SIMP scenario, we consider the two-component SIMP DM model (hereafter we dub it $a$SIMP model) studied in Ref.\,\cite{Ho:2022erb}, where the SM model is extended with a vector-like fermion, $N$, and three complex singlet scalars, $X, S$, and $\phi$.\,\,These exotic particles possess dark charges under a gauged U$(1)^{}_\tf{D}$ symmetry, and all SM particles are  dark neutral under this new symmetry.\,\,We summarize the particle contents with their charge assignments in Tab.\,\ref{tab:1}.\,\,In our setup, the $N$ and $X$ are selected as SIMP DM candidates, and the unstable particle $S$ bridges them.\,\,In particular, the $\phi$ particle develops a vacuum expectation value (VEV), which breaks the U$(1)^{}_\tf{D}$ symmetry.\,\,After the U$(1)^{}_\tf{D}$ symmetry breaking, these new particles can accidentally have a ${}^{}\Zf$ symmetry, stabilizing the DM particles in this model.

\begin{table}[t!]
\begin{center}
\def\arraystretch{1.3}
\begin{tabular}{|c||c||c|c|c|c|}
\hline
$\vphantom{|_|^|}$                            
& ~$H$~ & ~$N$~ & ~$X$~ & ~$S$~ & ~$\phi$~ 
\\\hline\hline 
~\,SU$(2)\vphantom{|_|^|}$~         
& ~$\mathbf{2}$~ & ~$\mathbf{1}$~ & ~$\mathbf{1}$~ & ~$\mathbf{1}$~ & ~$\mathbf{1}$~ 
\\\hline
~\,U$(1)^{}_\tf{Y}\vphantom{|_|^|}$~      
& ~$-{}^{}1/2$~ & ~$0$~ & ~$0$~ & ~$0$~ & ~$0$~ 
\\\hline
~\,U$(1)^{}_\tf{D}\vphantom{|_|^|}$~      
& ~$0$~ & ~$-{}^{}1/8$~ & ~$+{}^{}1/12$~ & ~$+{}^{}1/4$~ & ~$-{}^{}1/2$~ 
\\\hline
~$\Zf$~      
& ~$+{}^{}1$~ & ~$\pm{}^{}{}^{}i$~ & ~$-1$~ & ~$-1$~ & ~$+{}^{}1$~ 
\\\hline  
\end{tabular}
\caption{Charge assignments of the fermion and scalars in the $a$SIMP model, where 
$H$ is the SM Higgs doublet and $i =\sqrt{-1}$.}
\vs{-0.5cm}
\label{tab:1}
\end{center}
\end{table}

Since the particle contents and the Lagrangian density in the $a$SIMP model are exactly as same as the ones in Ref.\,\cite{Ho:2022erb}, for our purpose, here we only write down the relevant interactions and the mass spectra of the new particles in this model.\,\,

First, the Lagrangian density for the complex scalar fields in the $a$SIMP model is given by
\begin{eqnarray}
{\cal L}_\tf{scalar}
\,=\,
|{}^{}{\cal D}_\rho {}^{}H|^2 +
|{}^{}{\cal D}_\rho {}^{}X|^2 +
|{}^{}{\cal D}_\rho {}^{}S|^2 +
|{}^{}{\cal D}_\rho {}^{}{}^{}\phi {}^{}|^2 \,-
{\cal V}(H, X, S, \phi)
~,
\end{eqnarray}
where ${}^{}{\cal D}_\rho = \partial_\rho + (i/2){}^{} g^{}_\tf{W} \tau^a W^a_\rho + i g^{}_\tf{Y} {\cal Q}^{}_\tf{Y} B^{}_\rho + i g^{}_\tf{D} {\cal Q}^{}_\tf{D} C^{}_\rho{}^{}$ is the covariant derivative with $g^{}_\tf{W}{}^{}(W^a_\rho)$, $g^{}_\tf{Y}{}^{}(B^{}_\rho{}^{})$, and $g^{}_\tf{D}{}^{}(C^{}_\rho)$ being the SU$(2)$, U$(1)^{}_\tf{Y}$, and U$(1)^{}_\tf{D}$ gauge couplings (fields), respectively\,; $\tau^a\,\big(a = 1, 2, 3)$ the Pauli matrices, and ${\cal Q}^{}_\tf{Y}\,({\cal Q}^{}_\tf{D})$ the hypercharge (dark charge) operator.\,\,The scalar potential \,${\cal V} = {\cal V}(H, X, S, \phi)$\, is given by
\begin{eqnarray}\label{potential}
{\cal V} 
\,\Eq\,
\mu_h^2 {}^{} |H|^2 +
\mu_X^2 {}^{} |X|^2 +
\mu_S^2 {}^{} |S|^2 +
\mu_\phi^2 {}^{} | \phi |^2 
+
\lambda^{}_h |H|^4 +
\lambda^{}_X |X|^4 +
\lambda^{}_S |S|^4 +
\lambda^{}_\phi | \phi |^4 
\nn[0.1cm]
&&
+\,
\lambda_{h X} |H|^2 |X|^2 +
\lambda_{h S} |H|^2 |S|^2 +
\lambda_{h \phi} |H|^2 |\phi |^2
+
\lambda_{X \hs{-0.03cm} \phi} |X|^2 |\phi |^2 +
\lambda_{S \phi} |S|^2 |\phi |^2
\nn[0.1cm]
&&
+\,
\lambda_{X \hs{-0.03cm} S} |X|^2 |S|^2 + 
\sx{1.2}{\big(}{}^{}{}^{}
\lambda^{}_3 {}^{} X^3 \hs{-0.03cm} S^\ast +
\tfrac{1}{\sqrt2} {}^{}  \kappa {}^{} \upsilon^{}_\phi {}^{} S^2 \phi +
\text{h.c.}^{}
\sx{1.2}{\big)}
~,
\end{eqnarray}
where $\upsilon^{}_\phi$ is the VEV of $\phi$.\,\,The hermiticity of the scalar potential requires that the quadratic and quartic couplings except the $\lambda^{}_3$ and $\kappa$ must be real.\,\,However, we will take $\lambda^{}_3 > 0$ because one can redefine the $X$ field to absorb the phase of $\lambda^{}_3$.\,\,On the other hand, the role of the $\kappa$ coupling is to trigger the 
U$(1)^{}_\tf{D}$ symmetry breaking, and it is nothing to do with our numerical study.\,\,Thus, we assume that the $\kappa$ is nonzero but negligible.\,\,Also, we turn off the mass mixing between $H$ and $\phi$ for simplicity.\,\,Under these assumptions, the new scalar masses are given by
\begin{eqnarray}
m_X^2
\,=\,
\mu_X^2 
+ 
\tfrac{1}{2} 
\sx{1.1}{\big(} 
\lambda^{}_{h X} {}^{} \upsilon_h^2 +
\lambda^{}_{X \hs{-0.03cm} \phi} \upsilon_\phi^2 {}^{} 
\sx{1.1}{\big)} 
~,\quad
m_S^2 
\,=\,
\mu_S^2
+
\tfrac{1}{2}
\sx{1.1}{\big(} 
\lambda^{}_{h S}  {}^{} \upsilon_h^2 + \lambda^{}_{S \phi} \upsilon_\phi^2 {}^{} 
\sx{1.1}{\big)}
~,\quad
m_\phi^2 
\,=\, 2 {}^{} \lambda^{}_\phi \upsilon_\phi^2
~,
\label{Scalar_mass}
\end{eqnarray}
where $\upsilon^{}_h$ is the VEV of $H$.

Next, the Lagrangian density associated with the vector-like fermion is given by
\begin{eqnarray}\label{Yukawa}
{\cal L}^{}_N 
\,=\, 
\overline{N} \big({}^{}{}^{} i \gamma^\rho {}^{} {\cal D}_\rho - m^{}_N \big) N 
-
\tfrac{1}{2}
\sx{1.1}{\big(} 
\,y^{}_N \overline{N\raisebox{0.5pt}{$^\tf{c}$}} \hs{-0.03cm} N S + \tx{h.c.}
\sx{1.1}{\big)}
~,
\end{eqnarray}
where $m^{}_N$ is the Dirac mass of $N$, $y^{}_N$ is the Yukawa coupling, and $N^\tf{c}$ denotes the charge conjugation of $N$.\,\,Similar to the $\lambda^{}_3{}^{}$, we will take $y^{}_N > 0$ by absorbing its phase into the $N$ field or $S$ field.\,\,Note that the $S$ particle can decay into a pair of $\bar{N}$ if $m^{}_S > 2{}^{}m^{}_N$ or three $X$ particles if $m^{}_S > 3{}^{}m^{}_X$.\,\,Thus, although $S$ has a $\,\Zf$ charge, it cannot serve as a DM candidate if $m^{}_S > 2{}^{}m^{}_N$ or $3{}^{}m^{}_X$.

Finally, there is a new gauge interaction for the DM particles and SM fermions $f$ (with electric charge ${\cal Q}^{}_f$) mediated by a new massive gauge boson $Z'$.\,\,In the mass eigenbasis of the SM and new gauge bosons with $\epsilon \ll 1$ and $m^{}_Z \gg m^{}_{Z'}$, one can derive that
\begin{eqnarray}\label{Zprime}
{\cal L}^{}_{Z'}
\,=\,
-
\Big({}^{}
g^{}_\tf{D} {\cal Q}^{}_N \overline{N} \gamma^\rho N
+ i {}^{} g^{}_\tf{D} {\cal Q}^{}_X 
X^\ast \overleftrightarrow{\partial^\rho} X
+ g^{}_e {}^{}{\cal Q}^{}_f{}^{}{}^{} c^{}_\tf{w} {}^{} \epsilon {}^{}{}^{} \overline{f} {}^{}{}^{} \gamma^\rho f 
\Big)
Z'_\rho
~,
\end{eqnarray}
where ${\cal Q}^{}_N$ and ${\cal Q}^{}_X$ are dark charges of the $N$ and $X$ particles assigned in Tab.\,\ref{tab:1}, respectively, $g^{}_e = (4\pi\alpha)^{1/2}$ with $\alpha$ the fine structure constant, $c_\tf{w} = \cos\theta_\tf{w}$ with $\theta_\tf{w}$ the weak angle, and $\epsilon$ is the kinetic mixing strength of the U$(1)^{}_\tf{Y}$ and U$(1)^{}_\tf{D}$ gauge bosons.\,\,This new gauge interaction provides a vector portal coupling between the SM and dark sectors, which can thermalize the SIMP DM with the SM particles before the DM freeze-out (see Sec.\,\ref{sec:4}).\,\,Most importantly, we are able to probe the SIMP DM by future direct search experiments using electron target (see Sec.\,\ref{sec:6}).\,\,For more detailed discussions of the Lagrangian in the $a$SIMP model, see Ref.\,\cite{Ho:2022erb}.

\section{Annihilation cross sections in dark sector}\label{sec:3}

In this section, we will present the $3 \to 2$ and $2 \to 2$ annihilation cross sections in the dark sector.\,\,The calculations for these annihilation cross sections can be found in Ref.\,\cite{Ho:2022erb}.\,\,However, in Ref.\,\cite{Ho:2022erb}, we only consider the contributions of the two-loop diagrams to the $2 \to 2$ processes for certain reasons.\,\,In this work, we will incorporate the tree-level and one-loop contributions for the $2 \to 2$ processes to have a generic situation
and more accurate results. 

\begin{figure}[t!]
\hs{0.2cm}
\centering
\includegraphics[width=0.31\textwidth]{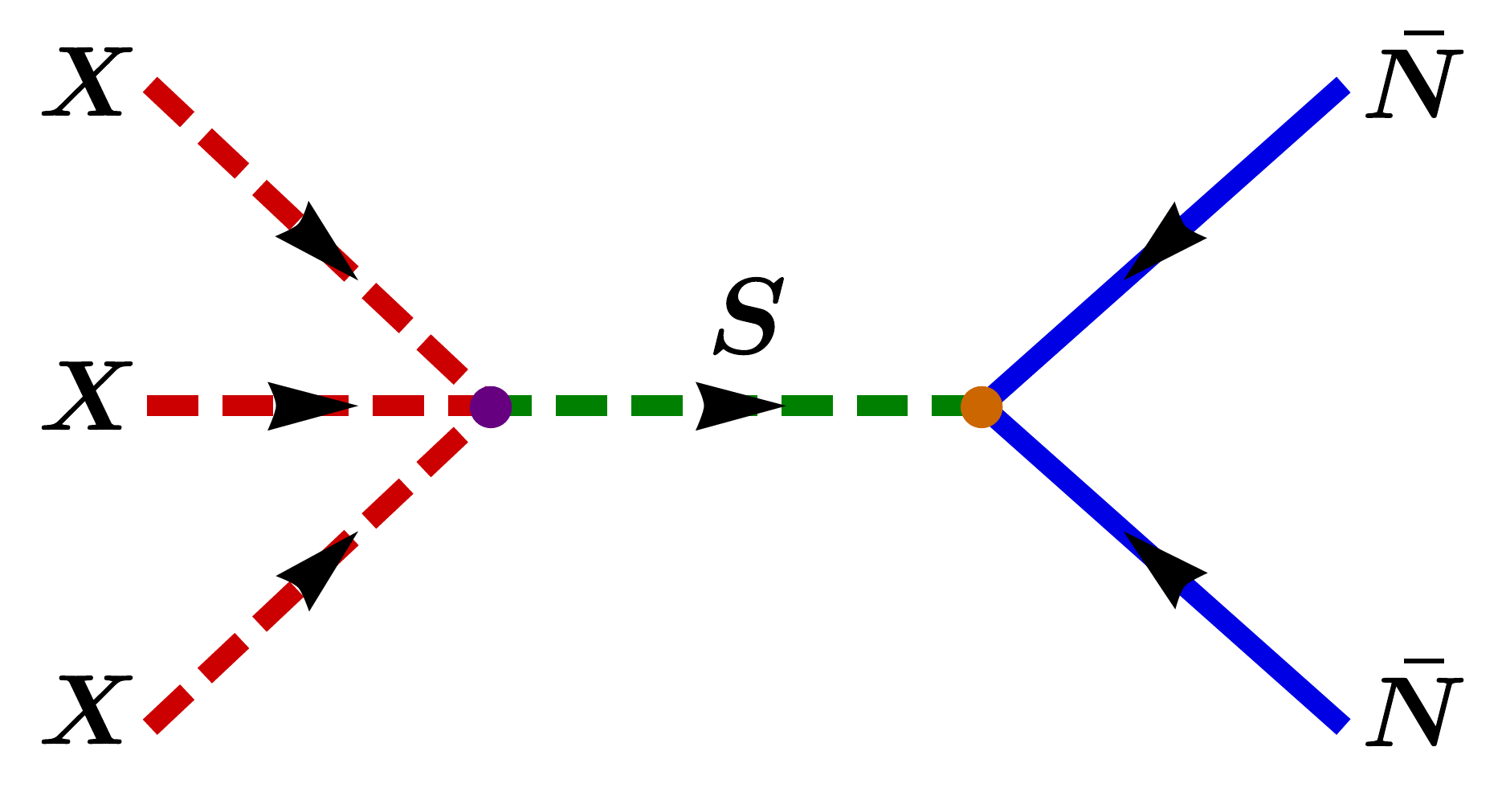}
\hs{0.2cm}
\includegraphics[width=0.31\textwidth]{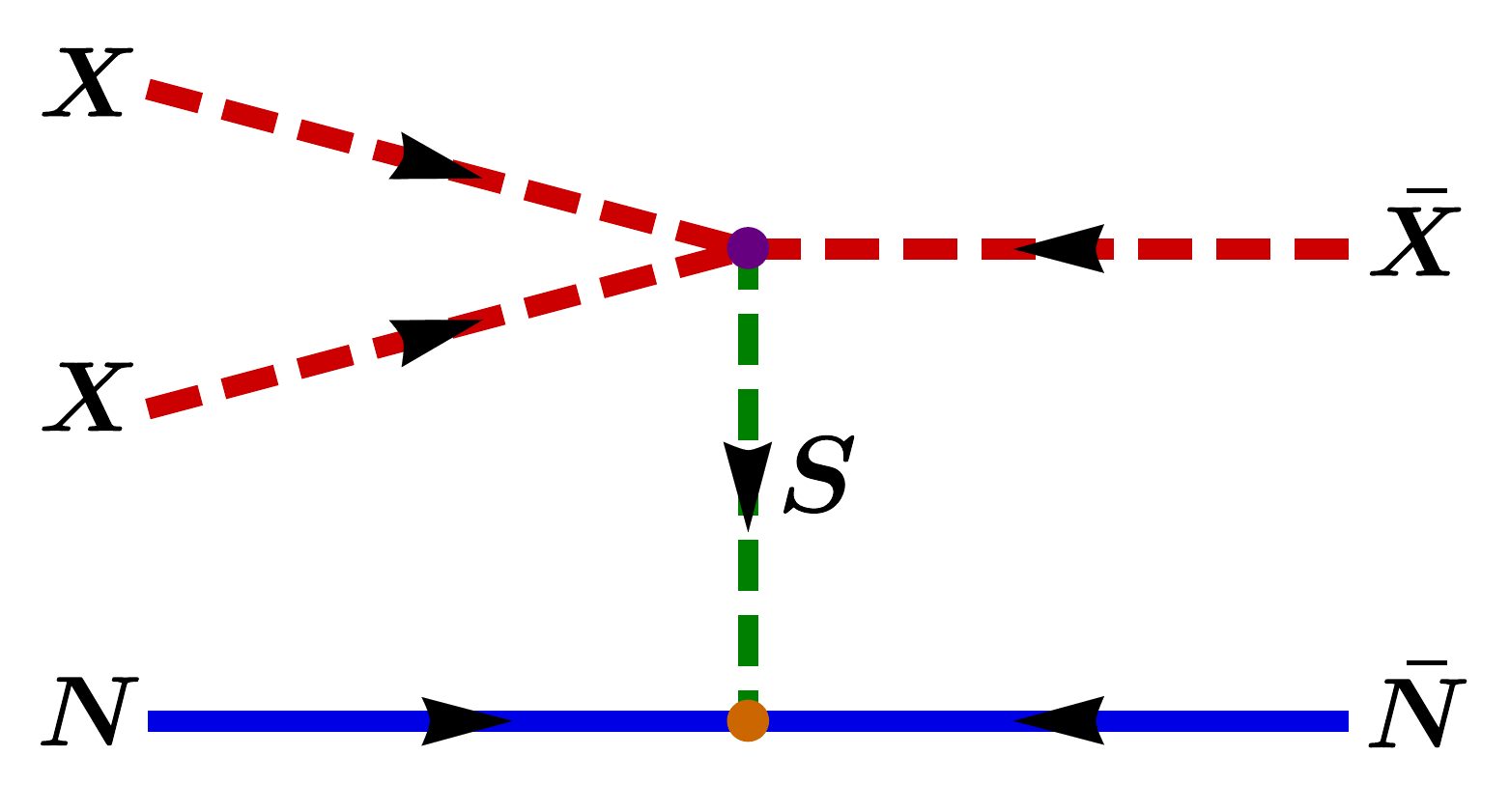}
\hs{0.2cm}
\includegraphics[width=0.31\textwidth]{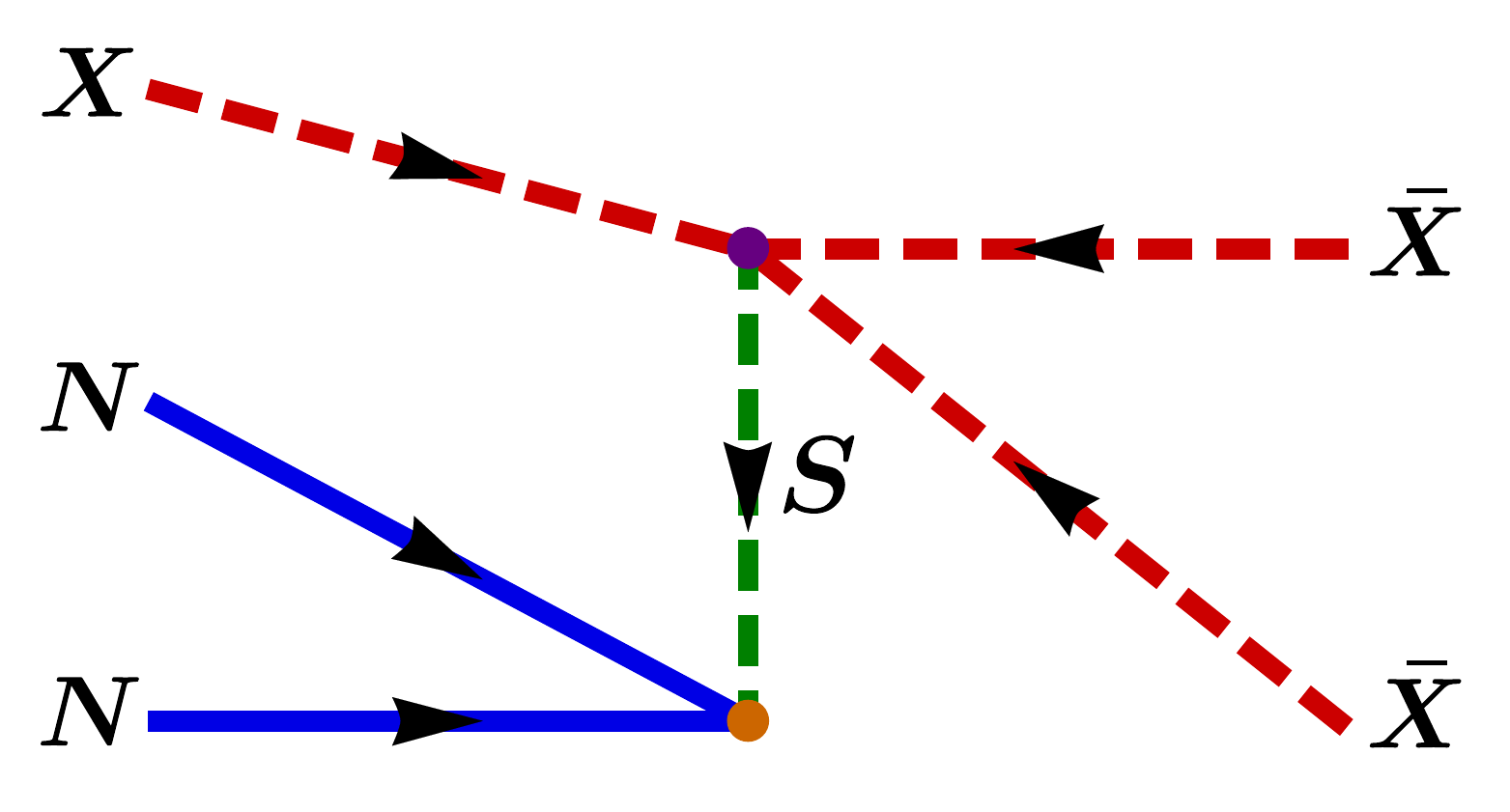}
\vs{-0.6cm}
\caption{Feynman diagrams of the $3 \to 2$ annihilation processes in the $a$SIMP model, where the arrows denote the direction of dark charge flow.\,\,The charge conjugation processes can be obtained by flipping the arrows of these diagrams.}
\label{fig:32ann}
\vs{0.2cm}
\end{figure}

Given the quartic interaction $X^3 S^\ast$ in Eq.\,\eqref{potential} and the Yukawa interaction $\overline{N\raisebox{0.5pt}{$^\tf{c}$}} \hs{-0.03cm} N S$ in Eq.\,\eqref{Yukawa}, all the possible 
$3 \to 2$ annihilation processes are drawn in Fig.\,\ref{fig:32ann}.\,\,For these $3 \to 2$ processes to be kinematically allowed, we will assume that the DM masses satisfy $3{}^{}m^{}_X \hs{-0.05cm} > \hs{-0.05cm} 2{}^{}m^{}_N \hs{-0.05cm} > \hs{-0.05cm} m^{}_X$.\,With this mass relation, the $2 \to 3$ and $2 \to 4$ processes such as $\NNXXX$ and $\XXNNNN$ are highly suppressed due to the Boltzmann tail at low temperatures.\,\,The thermally-averaged $3 \to 2$ annihilation cross sections first computed in Ref.\,\cite{Ho:2022erb} are
\begin{eqnarray}
\langle \sigma v^2 \rangle_{\hs{-0.03cm}\XXXNbNb}
\Eq
\langle \sigma v^2 \rangle_{\hs{-0.03cm}\XbXbXbNN}
\,=\,
\frac{x^3}{2} \hs{-0.03cm}
\int_{\hs{-0.03cm}0}^{\infty} \hs{-0.05cm}
\dd \beta \,(\sigma v^2)^\tf{BW}_{\hs{-0.03cm}\XXXNbNb} \,
\beta^2 e^{-{}^{}x{}^{}\beta}
\label{XXXNN}
~,
\\[0.1cm]
\langle \sigma v^2 \rangle_{\hs{-0.03cm}\XXNXbNb}
\Eq
\langle \sigma v^2 \rangle_{\hs{-0.03cm}\XbXbNbXN}
\nn[0.1cm]
\Eq
\frac{9\sqrt{3}\,\lambda^2_3 {}^{}{}^{} y^2_N}{32{}^{}\pi{}^{}m^5_X} 
\frac{
\big(1 + r^{}_N\big)
\sx{1.1}{\big[}\big(1+r^{}_N\big)\raisebox{1pt}{$^{\hs{-0.05cm}2}$} + r_N^2 \sx{1.1}{\big]} 
\sx{1.1}{\big[}\big(1+2{}^{}r^{}_N\big)\big(3 + 2{}^{}r^{}_N\big)\sx{1.1}{\big]}^{\hs{-0.05cm}1/2}}
{\big({}^{}2 + r^{}_N\big)\raisebox{1pt}{$^{\hs{-0.05cm}2}$} \sx{1.0}{\big[} r_S^2\big(1 + r^{}_N\big) + 
2{}^{}r^{}_N \sx{1.0}{\big]}\raisebox{1pt}{$^{\hs{-0.05cm}2}$}}
~,
\label{XXNXN}
\end{eqnarray}
where $x = m^{}_X/T$ is the dimensionless cosmic time variable with $T$ being the thermal plasma temperature, and $r^{}_{N,{}^{}S} \equiv m^{}_{N,{}^{}S}/m^{}_X$ with $3/2 > r^{}_N > 1/2$ and $r^{}_S > 2{}^{}r^{}_N$ based on the above assumption.\,\,Since the $\langle\sigma v^2 \rangle_{\hs{-0.03cm}\XNNXbXb} = \langle\sigma v^2 \rangle_{\hs{-0.03cm}\XbNbNbXX} = {\cal O}(x^{-1})$ are $p{}^{}{}^{}$-wave suppressed, thereby we do not include them in our numerical calculation.\,\,In order for the sizes of $\lambda^{}_3$ and $y^{}_N$ away from the perturbative bounds, we use the following Breit-Wigner cross section~\cite{Ho:2022erb,Ho:2017fte} in Eq.\,\eqref{XXXNN} 
\begin{eqnarray}
(\sigma v^2)^\tf{BW}_{\hs{-0.03cm}\XXXNbNb}
\Eq
\frac
{2{}^{}\pi \lambda_3^2{}^{}{}^{}r_S^2 \big({}^{}9 - 4{}^{} r_N^2\big)^{\hs{-0.05cm}3/2}}
{y_N^2{}^{}m_X^5 \big({}^{}r_S^2 - 4{}^{}r_N^2\big)\raisebox{-0.1pt}{$^{\hs{-0.03cm}3}$}}
\frac{\gamma^2_S}
{\big(\epsilon^{}_S - 2\beta/3 {}^{} \big)\raisebox{-0.1pt}{$^{\hs{-0.03cm}2}$} + \gamma_S^2}
~,
\end{eqnarray}
and consider the resonant mass region, where $m^{}_S \simeq 3{}^{}m^{}_X$ for our study.\,\,In this expression, $\epsilon^{}_S$ indicates the level of the resonant effect, and the $\gamma^{}_S$ is the normalized dimensionless width of the resonance, respectively of the forms as~\cite{Ho:2022erb,Ho:2017fte}
\begin{eqnarray}
\epsilon^{}_S
\,=\,
\frac{r_S^2}{9}-1
~,\quad
\gamma^{}_S
\,=\,
\frac{r_S^2{}^{}{}^{}y_N^2}{144{}^{}\pi}
\bigg(1-\frac{4{}^{}r_N^2}{r_S^2}\bigg)^{\hs{-0.15cm}3/2}
~.
\end{eqnarray}

\begin{figure}[t!]
\hs{0.2cm}
\centering
\includegraphics[width=0.34\textwidth]{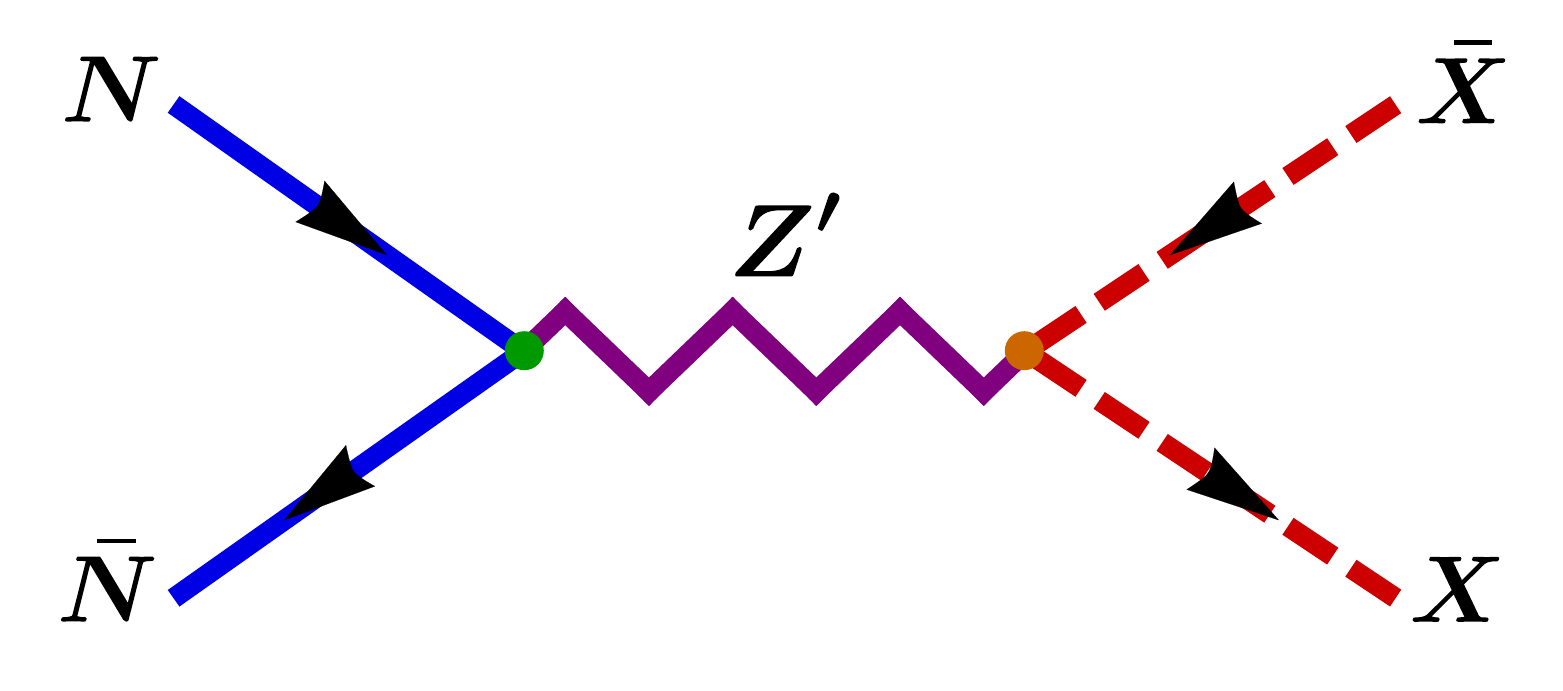}
\hs{0.3cm}
\includegraphics[width=0.41\textwidth]{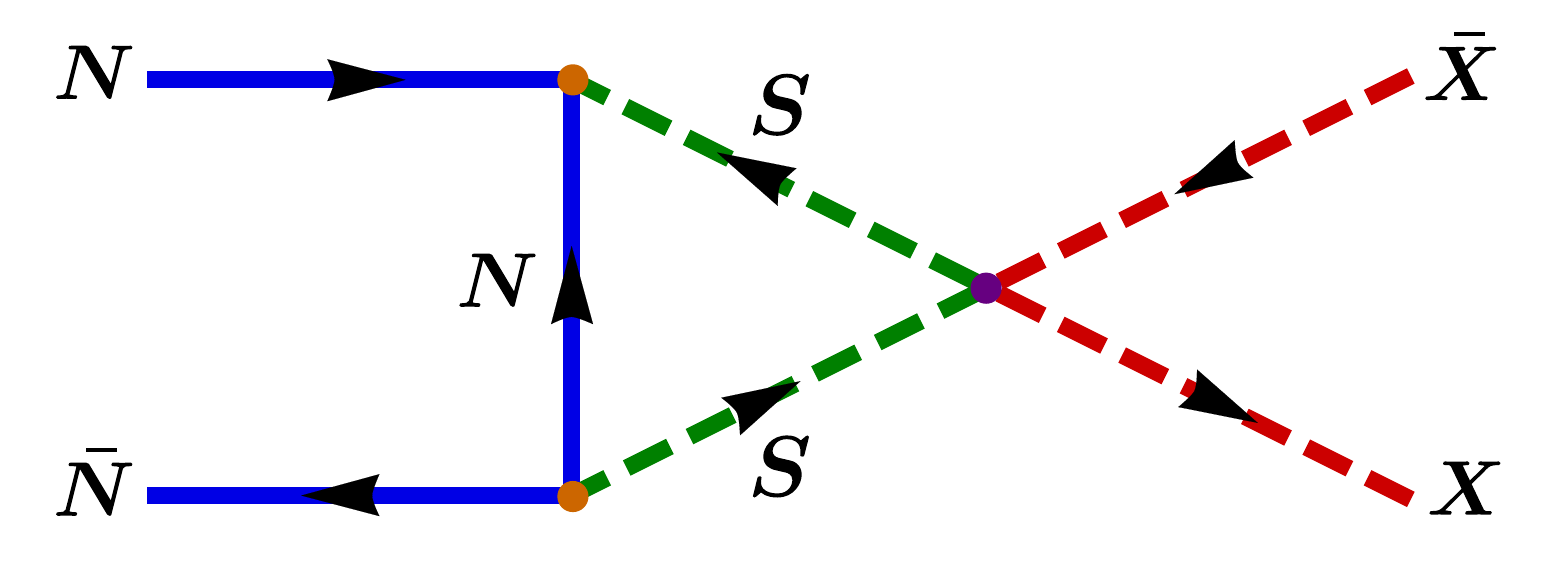}
\\[0.3cm]
\includegraphics[width=0.41\textwidth]{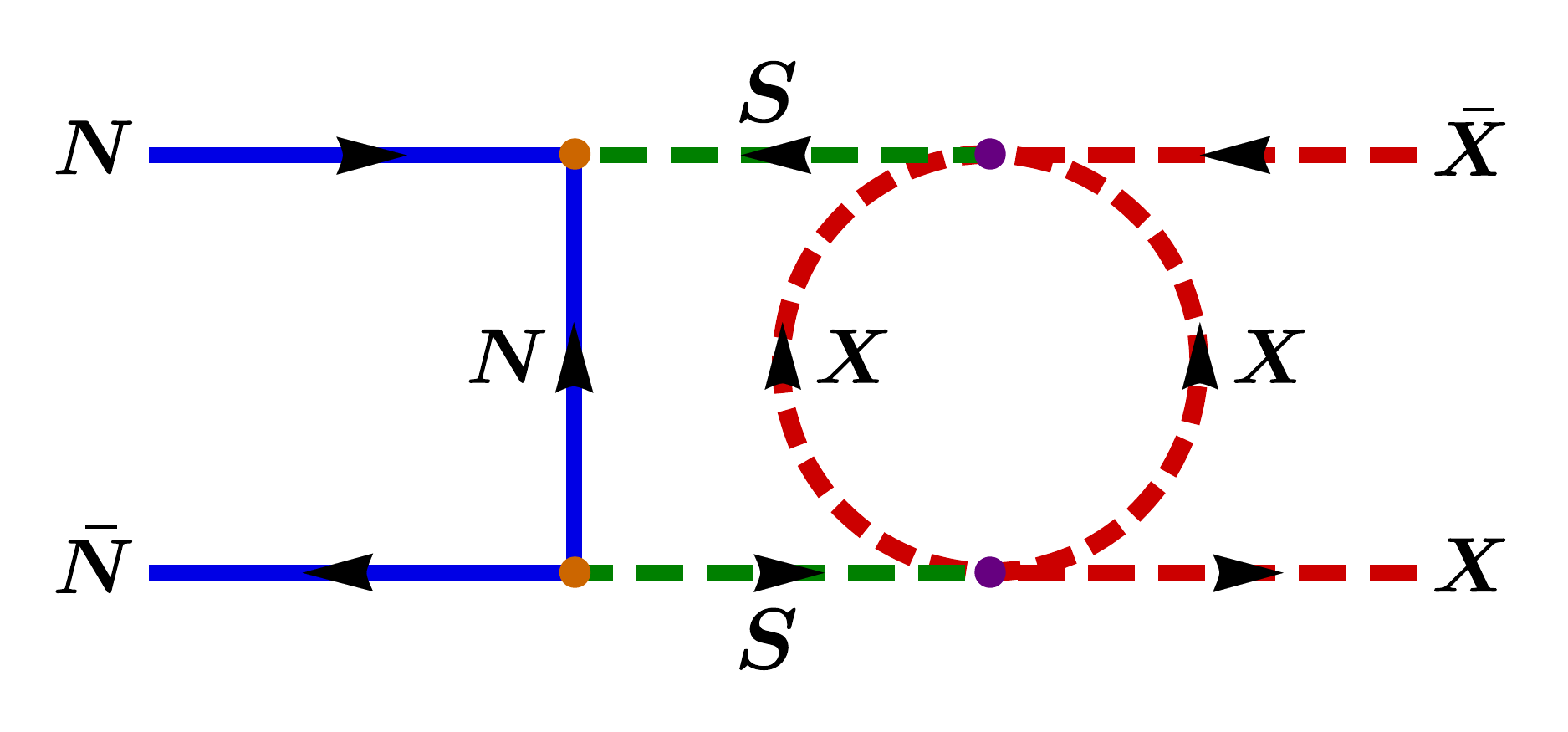}
\vs{-0.4cm}
\caption{Tree${}^{}{}^{}$-level, one${}^{}$-loop, and two${}^{}{}^{}$-loop Feynman diagrams for the $2 \to 2$ annihilation processes in the $a$SIMP model, where the Feynman diagrams for the inverse processes $\XXbNNb$ can be obtained by reversing the above diagrams.}
\label{fig:22ann}
\end{figure}

\begin{figure}[t!]
\hs{0.1cm}
\centering
\includegraphics[width=0.48\textwidth]{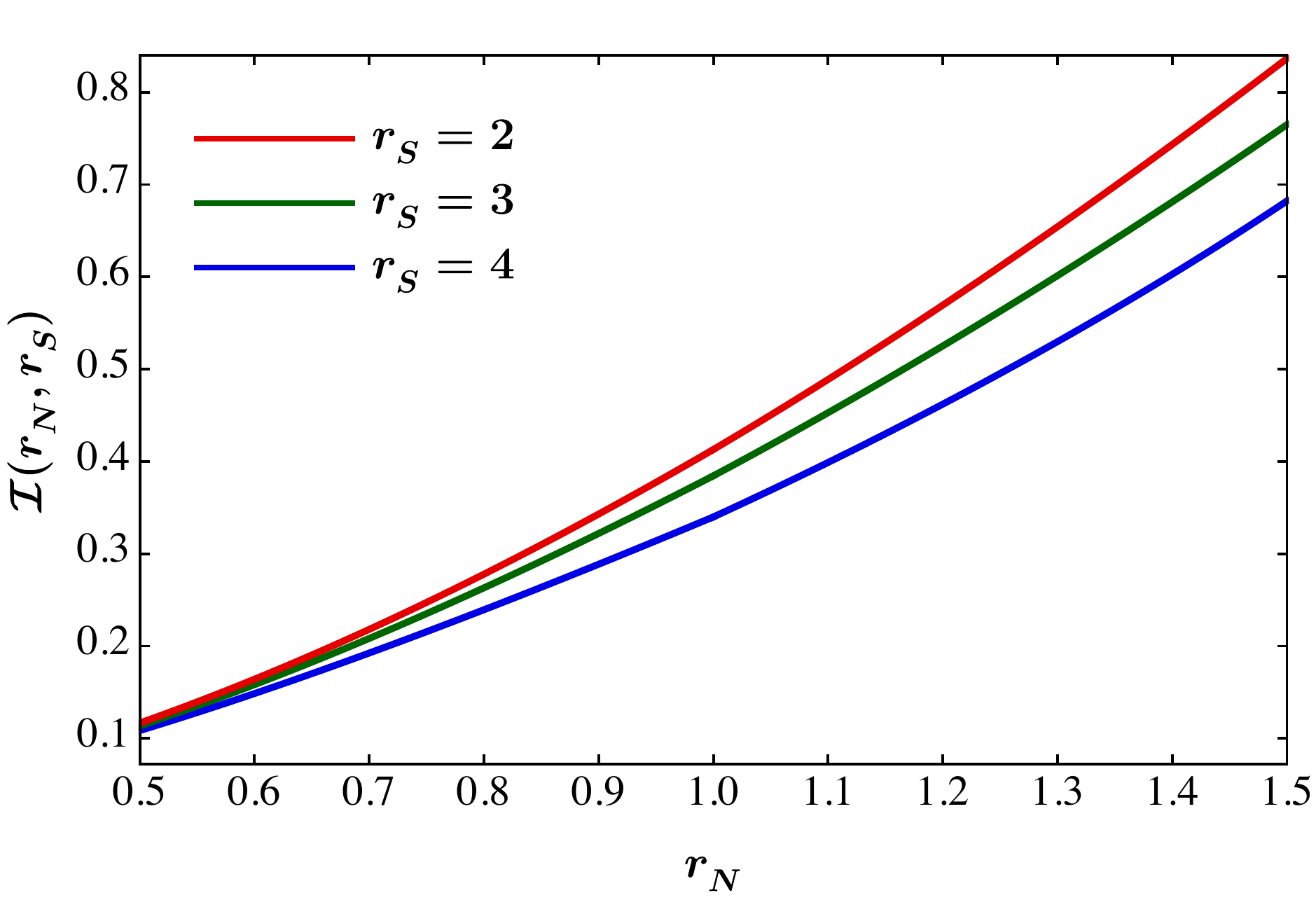}
\vs{-0.3cm}
\caption{One${}^{}{}^{}$-loop function $\,{\cal I}$ as a function of $\,r^{}_N$ with different choices of $r^{}_S{}^{}$.\,\,As indicated, the ${\cal I} \sim {\cal O}(0.1-1)$ in the DM mass range of interest.}
\label{fig:I1loop}
\end{figure}

\begin{figure}[t!]
\centering
\includegraphics[width=0.487\textwidth]{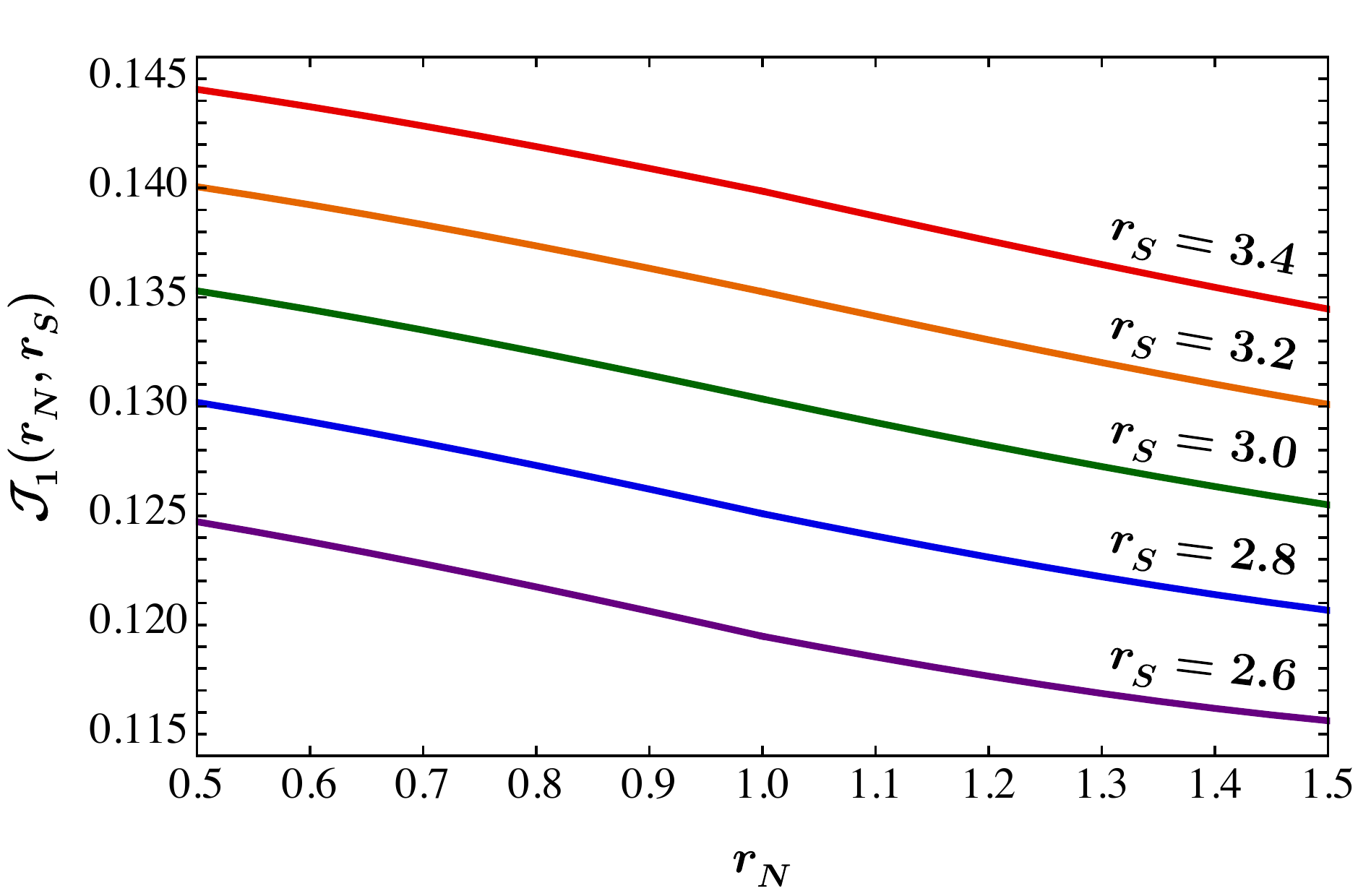}
\hs{0.1cm}
\includegraphics[width=0.48\textwidth]{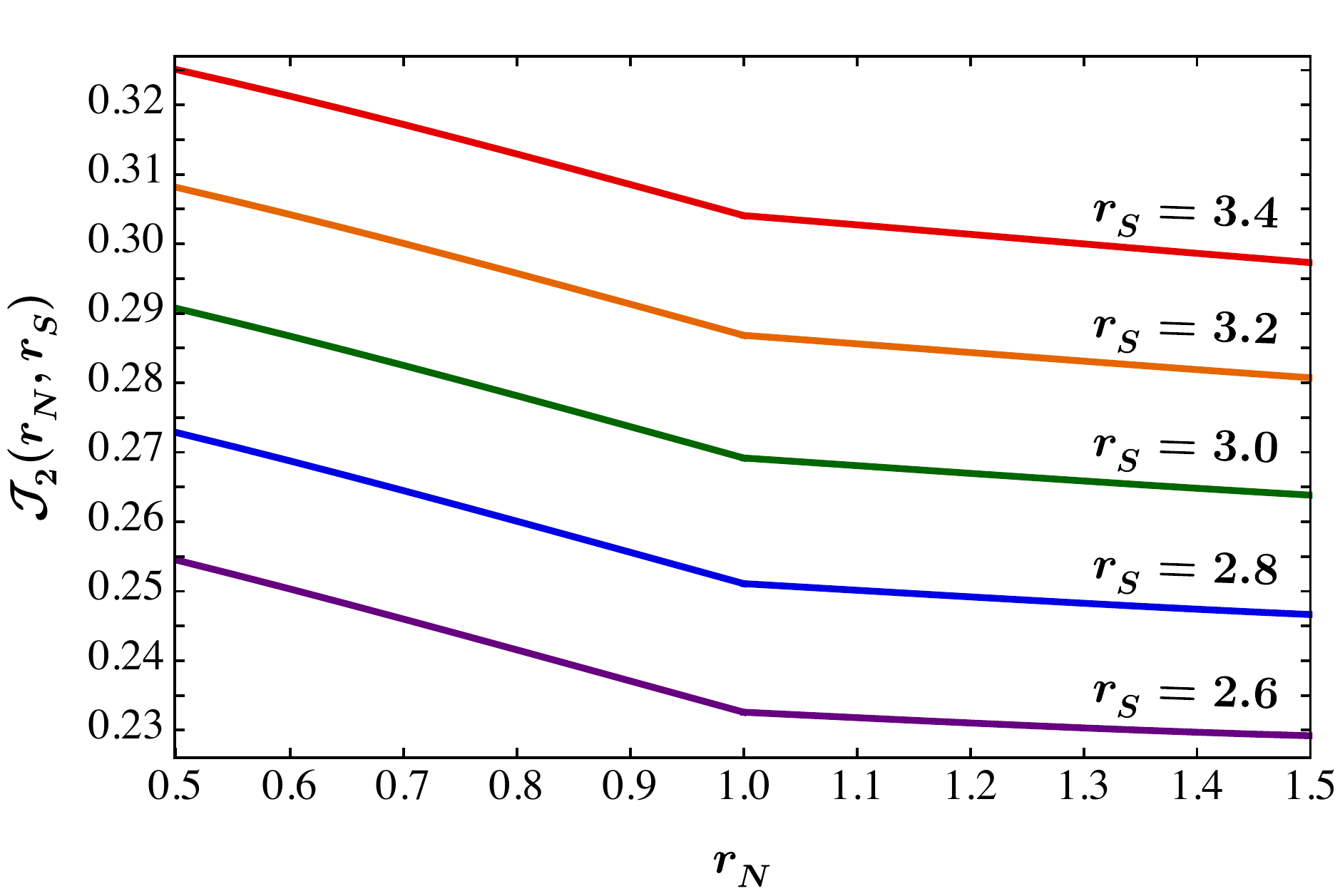}
\vs{-0.3cm}
\caption{Two${}^{}{}^{}$-loop functions ${\cal J}^{}_1$ and ${\cal J}^{}_2$ as functions of $\,r^{}_N$ with different values of $r^{}_S{}^{}$.\,\,As pointed out, the ${\cal J}^{}_{1,2} \sim {\cal O}(0.1)$ in the DM mass range of interest.}
\label{fig:J2loop}
\end{figure}

Next, with the dark gauge coupling $g^{}_\tf{D}$, the quartic couplings $\lambdaXS^{}$ and $\lambda^{}_3{}^{}$, and the Yukawa coupling $y^{}_N$, the $2 \to 2$ processes $\NNbXXb$ and $\XXbNNb$ are generated via the tree${}^{}{}^{}$-level, one${}^{}{}^{}$-loop, and two${}^{}{}^{}$-loop graphs as shown in Fig.\,\ref{fig:22ann}.\,\,The thermally-averaged $2 \to 2$ annihilation cross sections up to $p{}^{}{}^{}{}^{}$-wave contribution are computed as
\begin{eqnarray}
\langle \sigma v \rangle^{}_{\hs{-0.03cm}\NNbXXb}
\Eq
\frac{\big(r_N^2-1\big)^{\hs{-0.05cm}1/2}}{\pi{}^{}m_X^2{}^{}r^{}_N}
\sx{1.1}{\bigg[}
\big(r_N^2-1\big) {\cal Z}^{}_1
+
\frac{3}{2{}^{}x}
\bigg({}^{}{}^{}\frac{11-2{}^{}r_N^2}{6} {\cal Z}^{}_1 + {\cal Z}^{}_2 \bigg)
\sx{1.1}{\bigg]}
~,\quad
\label{22cs1}
\\[0.2cm]
\langle \sigma v \rangle^{}_{\hs{-0.03cm}\XXbNNb}
\Eq
\frac{\big(1-r_N^2\big)^{\hs{-0.05cm}1/2}}{\pi{}^{}m_X^2}
\sx{1.1}{\bigg[}
\big(1-r_N^2\big) {\cal Z}^{}_2
+
\frac{3}{2{}^{}x}
\bigg({}^{}{}^{}\frac{2+r_N^2}{3} {\cal Z}^{}_1 + \frac{5{}^{} r_N^2 -2}{2} {}^{} {\cal Z}^{}_2 \bigg)
\sx{1.1}{\bigg]}
~,
\label{22cs2}
\end{eqnarray}
where ${\cal Z}^{}_1$ and ${\cal Z}^{}_2$ are defined as
\begin{eqnarray}
{\cal Z}^{}_1
\,=\,
\sx{1.1}{\bigg[}{}^{}{}^{}
\frac{g_\tf{D}^2 {\cal Q}^{}_N {\cal Q}^{}_X}{2 \big(r_{Z'}^2 - 4{}^{}r_N^2\big)} 
+
\frac{9 {}^{} \lambda_3^2 {}^{}{}^{} y_N^2 {}^{}{}^{} {\cal J}^{}_1}{r_S^2 {}^{} (4\pi)^4}
\sx{1.1}{\bigg]}^{\hs{-0.07cm}2}
~,\quad
{\cal Z}^{}_2
\,=\,
\frac{r_N^2{}^{}y_N^4}{r_S^4{}^{}(4{}^{}\pi)^4}
\sx{1.1}{\bigg[}
\frac{\lambdaXS {}^{}{}^{}{}^{} {\cal I}}{4{}^{}r_N^2} 
+
\frac{9 {}^{} \lambda_3^2 {}^{}{}^{} {\cal J}^{}_2}{(4\pi)^2}
\sx{1.1}{\bigg]}^{\hs{-0.07cm}2}
\end{eqnarray}
with $r^{}_{Z'} = m^{}_{Z'}/m^{}_X$.\,\,In these definitions, ${}^{}{\cal I} = {\cal I}{}^{}(r^{}_N,r^{}_S)$ is a one-loop function, which is first derived in this work, in the form of a double integral as
\begin{eqnarray}
\hs{-0.4cm}
{\cal I}
\,=\,
\int_{\hs{-0.02cm}0}^1 \hs{-0.05cm} \dd w^{}_1
\int_{\hs{-0.02cm}0}^{1-w^{}_1} \hs{-0.05cm} \dd w^{}_2
\begin{cases}
\displaystyle\,
\frac{r_N^2{}^{}{}^{}r_S^2{}^{}(1-w^{}_1-w^{}_2)}{r_N^2\big[(1-w^{}_1-w^{}_2)^2 - 4{}^{}w^{}_1{}^{}w^{}_2\big] + r_S^2{}^{}(w^{}_1+w^{}_2)}
&\hs{-0.1cm} \tx{for} \,\,\, r^{}_N > 1
\\[0.5cm]
\displaystyle\,
\frac{r_N^2{}^{}{}^{}r_S^2{}^{}(1-w^{}_1-w^{}_2)}{r_N^2{}^{}(1-w^{}_1-w^{}_2)^2 + r_S^2{}^{}(w^{}_1+w^{}_2) - 4{}^{}w^{}_1{}^{}w^{}_2} 
&\hs{-0.1cm} \tx{for} \,\,\, r^{}_N < 1
\end{cases}
~,
\end{eqnarray}
and ${\cal J}^{}_{1,2} = {\cal J}^{}_{1,2}(r^{}_N,r^{}_S)$ are two-loop functions of the form in quintuple integrals as~\cite{Ho:2022erb}
\begin{eqnarray}\label{I1I2}
{\cal J}^{}_{1,2}
\,=\,
\int_{\hs{-0.02cm}0}^1 \hs{-0.05cm} \dd z^{}_1
\int_{\hs{-0.02cm}0}^1 \hs{-0.05cm} \dd z^{}_2
\int_{\hs{-0.02cm}0}^{1-z^{}_2} \hs{-0.05cm} \dd z^{}_3  
\int_{\hs{-0.02cm}0}^{{}^{}z^{}_1(1-z^{}_1)} \hs{-0.05cm} \dd z^{}_4
\int_{\hs{-0.02cm}0}^1 \hs{-0.05cm} \dd z^{}_5  
\,{}^{}{\cal K}^{}_{1,2}
\end{eqnarray}
with
\begin{eqnarray}
{\cal K}^{}_1
\,\Eq\,
\frac
{r_S^2 {}^{}{}^{} z_5^2 
\sx{1.1}{\big[} 2P^2 z_5^3 - 
\big(P^2 + 3{}^{}Q^2 \big) z_5^2 + 
\big(2{}^{}Q^2 + 3 \big) z^{}_5 - 2 {}^{} 
\sx{1.1}{\big]}}
{2 \big(P^2 z_5^2 - Q^2 z^{}_5 + 1 \big)\raisebox{1pt}{$^{\hs{-0.05cm}2}$}}
~,
\\[0.15cm]
{\cal K}^{}_2
\,\Eq\,
\frac{r_S^2 {}^{}{}^{} z_5^3 {}^{} \big(1-z^{}_2-z^{}_3{}^{}\big) \big(2P^2 z_5^2 - 3{}^{}Q^2 z^{}_5 + 3 \big)}
{2 \big(P^2 z_5^2 - Q^2 z_5 + 1\big)\raisebox{1pt}{$^{\hs{-0.05cm}2}$}}
~,
\end{eqnarray}
\vs{-0.3cm}
\begin{eqnarray}
P^2
\,\Eq\,
\begin{cases}
\,z^{}_4{}^{} 
\sx{1.2}{\big[} {}^{} 
r_N^2 \big(z^{}_2 - z^{}_3 + 1\big) \big(z^{}_2 - z^{}_3 - 1\big)+1
\sx{1.2}{\big]}
&\,\,\text{for} \,\,\, r^{}_N > 1
\\[0.3cm]
\,z^{}_4 {}^{} \sx{1.2}{\big[} {}^{} 
r_N^2 \big(z^{}_2 + z^{}_3 - 1\big)\raisebox{1pt}{$^{\hs{-0.03cm}2}$} - 
\big(2{}^{}z^{}_2 - 1\big)\big(2{}^{}z^{}_3 - 1\big) \sx{1.2}{\big]} 
&\,\,\text{for} \,\,\, r^{}_N < 1
\end{cases}
~,\quad
\label{P2}
\\[0.2cm]
Q^2
\,\Eq\,
\begin{cases}
\,1 + z^{}_4 {}^{} \sx{1.2}{\big[} 
2{}^{}r_N^2 \big(z^{}_2+z^{}_3-1\big) - r_S^2 \big(z^{}_2+z^{}_3{}^{}\big) + 1\sx{1.2}{\big]}
&\text{for}  \,\,\, r^{}_N > 1
\\[0.3cm]
\,1 + z^{}_4 {}^{} \sx{1.2}{\big[} \big(2 - r_S^2\big) \big(z^{}_2+z^{}_3{}^{}\big) -1 \sx{1.2}{\big]}
&\text{for}  \,\,\, r^{}_N < 1
\end{cases}
~.
\label{Q2}
\end{eqnarray}
We present the typical values of $\,{\cal I}$ and ${\cal J}^{}_{1,2}$ for $3/2 > r^{}_N > 1/2$ in Fig.\,\ref{fig:I1loop} and Fig.\,\ref{fig:J2loop}, respectively.\footnote{In Fig.\,3 of Ref.\,\cite{Ho:2022erb}, the solid (dashed) lines are only valid for $r^{}_N>1$ ($r^{}_N < 1$) based on Eqs.\,\eqref{P2} and \eqref{Q2} in this paper.\,\,We have made the corrections in these new figures.}\,\,Unlike the $3 \to 2$ cross sections, we include the $p{}^{}{}^{}{}^{}$-wave contribution for the $2 \to 2$ cross sections since it may be more dominant than the $s{}^{}{}^{}$-wave contribution when $r^{}_N \simeq 1$.\,\,One can also notice that the one${}^{}{}^{}$-loop (tree${}^{}{}^{}$-level) contribution is $p{}^{}{}^{}{}^{}$-wave suppressed for the $\NNbXXb$ ($\XXbNNb$) process.\,\,We will discuss their effects in Sec.\,\ref{sec:5}.

\section{Theoretical \& observational constraints}\label{sec:4}

In this section, we briefly summarize the constraints for the couplings and masses of the new particles in the $a$SIMP model as its particle contents and Lagrangian are as same as the ones in Ref.\,\cite{Ho:2022erb}.\,\,Besides, the CMB constraint which is missed in \cite{Ho:2022erb} will be discussed in this section.

From a theoretical perspective, the quartic, Yukawa, and dark gauge couplings should fulfill the perturbative conditions.\,\,We require that~\cite{Choi:2021yps,Perez:2021rbo,Allwicher:2021rtd}
\begin{eqnarray}\label{pertur}
\lambda^{}_{X,{}^{}S,{}^{}X \hs{-0.03cm} S,{}^{}3} < 4{}^{}\pi 
~,\quad
y^{}_N < \sqrt{8{}^{}\pi} 
~,\quad
g^{}_\tf{D}
< 4{}^{}\pi 
~,
\end{eqnarray}
as well as for the other quartic couplings.\,\,Besides, the unitarity of S-matrix sets a conservative bound for the scattering amplitude of self-interaction, where $|{\cal M}_\tf{self}| < 16{}^{}\pi$~\cite{Biswas:2021dan,Namjoo:2018oyn}, by which the quartic couplings $\lambda^{}_{X,{}^{}S} < 4{}^{}\pi$.\,\,On the other hand, the thermally-averaged annihilation cross sections are bounded from above by partial wave unitarity~\cite{Namjoo:2018oyn}.\,\,However, it places no stringent restrictions on the couplings and masses in this model.\,\,Moreover, the scalar potential at large values of the scalar fields should be bounded from below to stabilize the vacuum, for which the quartic couplings in the dark sector have to satisfy some relations.\,\,We find that~\cite{Choi:2016tkj,Ho:2022erb}
\begin{eqnarray}
\hs{-0.2cm}
\lambda^{}_{X,{}^{}{}^{}S} > 0
~,\quad
\lambda^{}_{X\hs{-0.03cm}S} + 2 \sqrt{\lambda^{}_X \lambda^{}_S} {}^{} > 0
~,\quad
|\lambda^{}_3| <
\sqrt{
\frac{
\big(12{}^{}\lambda^{}_X \lambda^{}_S + \lambda^{2}_{X\hs{-0.03cm}S}\big)\raisebox{1pt}{$\hs{-0.05cm}^{3/2}$} +
36{}^{}\lambda^{}_X \lambda^{}_S \lambda^{}_{X\hs{-0.03cm}S} -
\lambdaXS^3}
{54{}^{}\lambda^{}_S}}
~,
\end{eqnarray}
here we have assumed that the other quartic couplings associated with the new particles are positive but sufficiently small.\,\,For $\lambdaXS=0$, the above conditions are reduced to $\lambda_{X,{}^{}{}^{}S} > 0$ and $|\lambda^{}_3| < \big(16{}^{}\lambda^3_X \lambda^{}_S/27 \big)\raisebox{1pt}{$^{\hs{-0.05cm}1/4}$}$.\,\,Note that these conditions also guarantee that $\langle 0| X |0 \rangle = \langle 0| S |0 \rangle = 0$.

To avoid the temperature increase of SIMP DM due to the $3 \to 2$ annihilations, the SIMP DM  should keep thermal equilibrium with the SM particles at least before the DM freeze-out temperature, known as the SIMP condition~\cite{Hochberg:2015vrg,Hochberg:2018rjs}.\,\,As mentioned in Sec.\,\ref{sec:1}, the SIMP particles in this model naturally couple to the SM fermions via the new gauge interaction, which sets a lower bound on the product of the dark gauge coupling and kinetic mixing parameter.\,\,Referring to the detailed calculation in Ref.\,\cite{Ho:2022erb}, in the case of the nondegenerate DM masses, we demand that
\begin{eqnarray}\label{gDlower}
g^{}_\tf{D} \epsilon
\,\gtrsim\,
\frac{2 \times 10^{-4}}{\sqrt{{\cal Q}_N^2/r^{}_N + {\cal Q}_X^2}}
\bigg(\frac{g^{}_{\star,f}}{10.75}\bigg)^{\hs{-0.17cm}1/4}
\bigg(\frac{x^{}_f}{20}\bigg)^{\hs{-0.15cm}3}
\bigg(\frac{m^{}_{Z'}}{250\,\text{MeV}}\bigg)^{\hs{-0.15cm}2}
\bigg(\frac{m^{}_X}{20\,\text{MeV}}\bigg)^{\hs{-0.17cm}-3/2} 
~,
\end{eqnarray}
where $g^{}_{\star,f}$ is the effective energy degrees of freedom of the SM thermal bath at the freeze-out temparature of DM, $x^{}_f{}^{}$.\,\,In our numerical calculation, we will adopt the marginal values of $g^{}_\tf{D} \epsilon$ in Eq.\,\eqref{gDlower}.\,${}^{}$With such $g^{}_\tf{D} \epsilon$ values, the WIMP annihilation processes $\XX \to e^+e^-$ and $\NN \to e^+e^-$ are suppressed,\footnote{We have numerically checked that even with slightly larger $g^{}_\tf{D} \epsilon$ values, the WIMP annihilation processes only affect the predicted DM abundance by less than 1\%.} thus we can achieve the $a$SIMP scenario.

In cosmological observations, the low mass DM would contribute to the effective number of neutrino species, $N^\nu_\tf{eff}{}^{}$.\,\,The latest measurement from the Planck satellite gives $N^{\nu}_\tf{eff} = 2.99^{+0.34}_{-0.33}$ (95\% C.L.)~\cite{Aghanim:2018eyx}, which suggests that
the DM mass should be larger than 10\,MeV~\cite{Ho:2022erb,Smirnov:2020zwf}.\,\,To be conservative, we will assume that the DM masses $m^{}_{N,{}^{}X} \gtrsim 15\,\tx{MeV}$ in our model.\,\,On the other hand, the Planck collaboration also accurately measures the current abundance of DM, which shows that $\Omega^{}_\tf{DM} h^2 = 0.12 \pm 0.0012$ with $h$ the normalized Hubble constant~\cite{Aghanim:2018eyx}.\,\,On top of that, the annihilation cross section of a DM pair into the SM fermions may affect the CMB temperature and polarizations, which imposes a lower bound on the mass of DM.\,\,Quoting the analysis in Ref.\,\cite{Padmanabhan:2005es}, for a single symmetric DM species, we have
\begin{eqnarray}\label{mDMCMB}
m^{}_\tf{DM} 
\,\gtrsim\, 
(10-100)\,\tx{GeV} 
\sx{1.1}{\bigg(}
\frac{\langle \sigma v\rangle_{\tf{DM} + \overline{\tf{DM}} \to \tf{SM} + \overline{\tf{SM}}}}
{2 \times 10^{-26}\,\tx{cm}^3\,\tx{s}^{-1}}
\sx{1.1}{\bigg)}
~.
\end{eqnarray}
Accordingly, the light DM may suffer from the CMB constraint if the DM annihilation cross section
does not get suppression at the CMB temperature.\,\,However, as we shall see soon, this strict constraint can be escaped if the DM is extremely asymmetric.\,\,This is easy to understand since in this case, the DM is hard to find its anti-partner to annihilate into the SM particles.

Experimentally, there are several constraints for the kinetic mixing parameter, depending on the mass of the new gauge boson.\,\,In this model, the $Z'$ mainly decays into invisible particles, $Z' \to \XX,\NN$, and $S\hs{-0.01cm}\bar{S}$ since $g_\tf{D}^2{\cal Q}_j^2 \gg g_e^2 {}^{} \epsilon^2$.\,\,Also, we will focus on the $Z'$ with a few hundred MeV mass, where the measurements from the BaBar collaboration cap $\epsilon \lesssim 10^{-3}$~\cite{BaBar:2017tiz,Fabbrichesi:2020wbt}.

\section{Relic abundance and bouncing effect of DM}\label{sec:5}

In contrast to the WIMP and ADM, there is no approximate analytical solution for the relic density of asymmetric SIMP DM.\,\,To evaluate the relic density of DM in the $a$SIMP model, we have to numerically solve the coupled Boltzmann equations of the comoving number yields $Y^{}_{N,\bar{N}}$ and $Y^{}_{X,\bar{X}}$.\,Using the formula in Ref.\,\cite{Ho:2021ojb}, the Boltzmann equations are written as follows
\begin{eqnarray}
\frac{\dd \YN^{}}{\dd x} 
\Eq
-\frac{s(x)^2}{x H(x)}
\nn[0.1cm]
&&
\sx{1.1}{\Bigg\{} 
2{}^{}\langle \sigma v^2 \rangle^{}_{\hs{-0.03cm}\XXNXbNb} \hs{-0.03cm}
\sx{1.2}{\bigg[} 
\YX^2 \YN^{} - \YXb^{} \YNb^{} \frac{(\YeqX)^2 \YeqN}{\YeqXb \YeqNb} 
\sx{1.2}{\bigg]}
+
2{}^{}\langle \sigma v^2 \rangle^{}_{\hs{-0.03cm}\XNNXbXb} \hs{-0.03cm}
\sx{1.2}{\bigg[} 
\YX^{} \YN^2 - \YXb^2 \frac{\YeqX (\YeqN)^2}{(\YeqXb)^2} 
\sx{1.2}{\bigg]}
\nn[0.1cm]
&&\hs{0.4cm}{-}\,
8{}^{}\langle \sigma v^2 \rangle^{}_{\hs{-0.03cm}\XbXbXbNN} \hs{-0.03cm}
\sx{1.2}{\bigg[} 
\YXb^3 - \YN^2 \frac{(\YeqXb)^3}{(\YeqN)^2} 
\sx{1.2}{\bigg]} \hs{-0.05cm}
-
2{}^{}\langle \sigma v^2 \rangle^{}_{\hs{-0.03cm}\XbXbNbXN} \hs{-0.03cm}
\sx{1.2}{\bigg[} 
\YXb^2 \YNb^{} - \YX^{} \YN^{} \frac{(\YeqXb)^2 \YeqNb}{\YeqX \YeqN} 
\sx{1.2}{\bigg]}
\sx{1.1}{\Bigg\}}
\nn[0.1cm]
&&
-\frac{s(x)}{x H(x)}
\nn[0.1cm]
&&
\sx{1.1}{\Bigg\{}
\langle \sigma v \rangle^{}_{\hs{-0.03cm}\NNbXXb} \hs{-0.03cm}
\sx{1.2}{\bigg[} 
\YN^{} \YNb^{} - \YX^{} \YXb^{} \frac{\YeqN \YeqNb}{\YeqX \YeqXb} 
\sx{1.2}{\bigg]} \hs{-0.05cm}
-
4{}^{}\langle \sigma v \rangle^{}_{\hs{-0.03cm}\XXbNNb} \hs{-0.03cm}
\sx{1.2}{\bigg[} 
\YX^{} \YXb^{} - \YN^{} \YNb^{} \frac{\YeqX \YeqXb}{\YeqN \YeqNb} 
\sx{1.2}{\bigg]} 
\sx{1.1}{\Bigg\}}
~,
\label{dYNdx}
\end{eqnarray}
\vs{-0.5cm}
\begin{eqnarray}
\frac{\dd \YX^{}}{\dd x} 
\Eq
-\frac{s(x)^2}{x H(x)}
\nn[0.1cm]
&&
\sx{1.1}{\Bigg\{}
12{}^{}\langle \sigma v^2 \rangle^{}_{\hs{-0.03cm}\XXXNbNb} \hs{-0.03cm}
\sx{1.2}{\bigg[} 
\YX^3 - \YNb^2 \frac{(\YeqX)^3}{(\YeqNb)^2} 
\sx{1.2}{\bigg]}
+
4{}^{}\langle\sigma v^2 \rangle^{}_{\hs{-0.03cm}\XXNXbNb} \hs{-0.03cm}
\sx{1.2}{\bigg[} 
\YX^2 \YN^{} - \YXb^{} \YNb^{} \frac{(\YeqX)^2 \YeqN}{\YeqXb \YeqNb} 
\sx{1.2}{\bigg]}
\nn[0.1cm]
&&\hs{0.4cm}{+}\,
\langle \sigma v^2 \rangle^{}_{\hs{-0.03cm}\XNNXbXb} \hs{-0.03cm}
\sx{1.2}{\bigg[} 
\YX^{} \YN^2 - \YXb^2 \frac{\YeqX (\YeqN)^2}{(\YeqXb)^2} 
\sx{1.2}{\bigg]} \hs{-0.05cm}
-
2{}^{}\langle \sigma v^2 \rangle^{}_{\hs{-0.03cm}\XbXbNbXN}  \hs{-0.03cm}
\sx{1.2}{\bigg[} 
\YXb^2 \YNb^{} - \YX^{} \YN^{} \frac{(\YeqXb)^2 \YeqNb}{\YeqX \YeqN} 
\sx{1.2}{\bigg]}
\nn[0.1cm]
&&\hs{0.4cm}{-}\,
2{}^{}\langle \sigma v^2 \rangle^{}_{\hs{-0.03cm}\XbNbNbXX} \hs{-0.03cm}
\sx{1.2}{\bigg[} 
\YXb^{} \YNb^2 - \YX^2 \frac{\YeqXb (\YeqNb)^2}{(\YeqX)^2} 
\sx{1.2}{\bigg]} \hs{-0.05cm}
\Bigg\}
\nn[0.1cm]
&&
-\frac{s(x)}{x H(x)}
\nn[0.1cm]
&&
\sx{1.1}{\Bigg\{}
4{}^{}\langle \sigma v \rangle^{}_{ \hs{-0.03cm}\XXbNNb}
\sx{1.2}{\bigg[} 
\YX^{} \YXb^{} - \YN^{} \YNb^{} \frac{\YeqX \YeqXb}{\YeqN \YeqNb} 
\sx{1.2}{\bigg]} 
-
\langle \sigma v \rangle^{}_{\hs{-0.03cm}\NNbXXb} \hs{-0.03cm}
\sx{1.2}{\bigg[} 
\YN^{} \YNb^{} - \YX^{} \YXb^{} \frac{\YeqN \YeqNb}{\YeqX \YeqXb} 
\sx{1.2}{\bigg]} \hs{-0.05cm}
\sx{1.1}{\Bigg\}}
~,
\label{dYXdx}
\end{eqnarray}
and $\dd \YNb^{}/\dd x$ and $\dd \YXb^{}/\dd x$ can be obtained by changing the right-hand sides of \eqref{dYNdx} and \eqref{dYXdx} with $X \leftrightarrow \bar{X}$ and $N \leftrightarrow \bar{N}$, respectively, where $Y^0_j$ is the (zero chemical potential) equilibrium comoving number yield of the DM species $j$ (with the number of spin states $g^{}_j$), of the form~\cite{Gondolo:1990dk}
\begin{eqnarray}
Y^0_{j}
\,=\,
\frac{45}{4{}^{}\pi^4}
\frac{g^{}_j}{g^{}_{\star s}(x)} 
\big(r^{}_j {}^{} x\big)^{\hs{-0.05cm}2}
K^{}_2 \hs{-0.03cm} \big(r^{}_j {}^{} x\big)
\end{eqnarray}
with $K^{}_2(x)$ being the modified Bessel function of the second kind.\,\,The $s(x)$ and $H(x)$ are the comoving entropy density and the Hubble parameter, respectively, which are expressed as
\begin{eqnarray}
s(x) \,=\, 
\frac{2{}^{}\pi^2 g^{}_{\star s}(x)}{45} \frac{m_X^3}{x^3} ~,\quad
H(x) \,=\, 
\sqrt{\frac{\pi^2 g^{}_{\star}(x)}{90}} \frac{m_X^2}{x^2 m^{}_\tf{Pl}} 
\end{eqnarray}
with $g^{}_{\star s}(x)$ being the effective entropic degrees of freedom of the SM thermal plasma~\cite{Saikawa:2018rcs}, and $m^{}_\tf{Pl} = 2.4 \times 10^{18}\,\tx{GeV}$ the reduced Planck mass.\,\,Note that the DM comoving number density $Y^\tf{eq}_j$ during the chemical equilibrium is not equal to $Y^0_j$ but $Y^\tf{eq}_j = Y^0_j e^{{}^{}\mu_j/T}$ due to the nonzero chemical potential for asymmetric DM.\,\,However, we still use $Y^0_j$ in the Boltzmann equations because the chemical potentials are cancelled in the ratios, e.g.\,\,$(Y^\tf{eq}_{\white{\bar{\black{X}}}})^3/(Y^\tf{eq}_{\bar{N}})^2 = (\YeqX)^3/(\YeqNb)^2$ as well as for the others.\,\,Now, doing proper addition, subtraction, and multiplication for the Boltzmann equations, one can find that
\begin{eqnarray}
3\bigg(\frac{\dd \YNb^{}}{\dd x} - \frac{\dd \YN^{}}{\dd x}\bigg) +
2\bigg(\frac{\dd \YX^{}}{\dd x} - \frac{\dd \YXb^{}}{\dd x}\bigg) 
\,=\,0
~.
\end{eqnarray}
This implies that we can define a conserved quantity, dark asymmetry
\begin{eqnarray}\label{etaDM}
\etaDM
\,=\, 
3{}^{}\etaN +
2{}^{}\etaX  
\quad \tx{with} \quad
\etaN \,=\, \YNb^{} - \YN^{} ~,\quad
\etaX \,=\, \YX^{} - \YXb^{} ~,\quad
\end{eqnarray}
which is a constant all the cosmic time.\,\,On the other hand, when the DM particles are in the chemical equilibrium, their chemical potentials satisfy the following relations
\begin{eqnarray}\label{murelation}
\mu(x) \,\equiv\, \muX = -{}^{}\muXb ~,\quad
\muN = -{}^{}\muNb ~,\quad
3{}^{}\muX = 2{}^{}\muNb ~,\quad
3{}^{}\muXb = 2{}^{}\muN ~,\quad
\end{eqnarray}
for the $x$ before the time of DM chemical freeze-out.\,\,Using Eq.\,\eqref{murelation} and the dark asymmetry in the chemical equilibrium period, $\etaDM = 3{}^{}\big({}^{}{}^{}\YNb^\tf{eq} - \YN^\tf{eq}\big) + 2{}^{}\big({}^{}{}^{}\YX^\tf{eq} - \YXb^\tf{eq}\big)$, one can show that
\begin{eqnarray}
\etaDM
\Eq
6{}^{}Y^0_N(x) \sinh\hs{-0.1cm} \sx{1.1}{\bigg[}\frac{3\mu(x)}{2{}^{}m^{}_X}{}^{}x\sx{1.1}{\bigg]} +
4{}^{}Y^0_X(x) \sinh\hs{-0.1cm} \sx{1.1}{\bigg[}\frac{\mu(x)}{m^{}_X}{}^{}x\sx{1.1}{\bigg]} 
~,
\end{eqnarray}
by which the $\mu(x)$ can be solved numerically for given $\etaDM$ and DM masses.\,\,Then, by setting appropriate initial conditions $Y^{}_j(x_\tf{ini.}\hs{-0.03cm}) = Y^\tf{eq}_j(x_\tf{ini.}\hs{-0.03cm})$ with $10 < x_\tf{ini.} \hs{-0.05cm} < 20$, we can numerically solve the Boltzmann equations to obtain the $Y^{}_j(x)$, and estimate the DM relic density as~\cite{Bhattacharya:2019mmy}
\begin{eqnarray}
\Omega^{}_\tf{DM} h^2 
\,=\,
\sum_{j{}^{}={}^{}\white{\bar{\black{N}}},\bar{N},\white{\bar{\black{X}}},\bar{X}} 
\Omega_j {}^{} h^2 
\,\simeq\,
2.745 \times 10^5 {}^{}
\sx{0.9}{\bigg(}
\frac{m^{}_X}{\text{MeV}}
\sx{0.9}{\bigg)}
\Big[{}^{}{}^{}
Y^\infty_{\white{\bar{\black{X}}}} + Y^\infty_{\bar{X}} 
+ 
r^{} _N \big({}^{}{}^{}Y^\infty_{\white{\bar{\black{N}}}} + Y^\infty_{\bar{N}} \big)
\Big]
~,
\end{eqnarray}
where $Y^\infty_j = Y^{}_j(x \to \infty)$ is the present comoving number density of DM.

\begin{figure}[t!]
\centering
\includegraphics[width=0.49\textwidth]{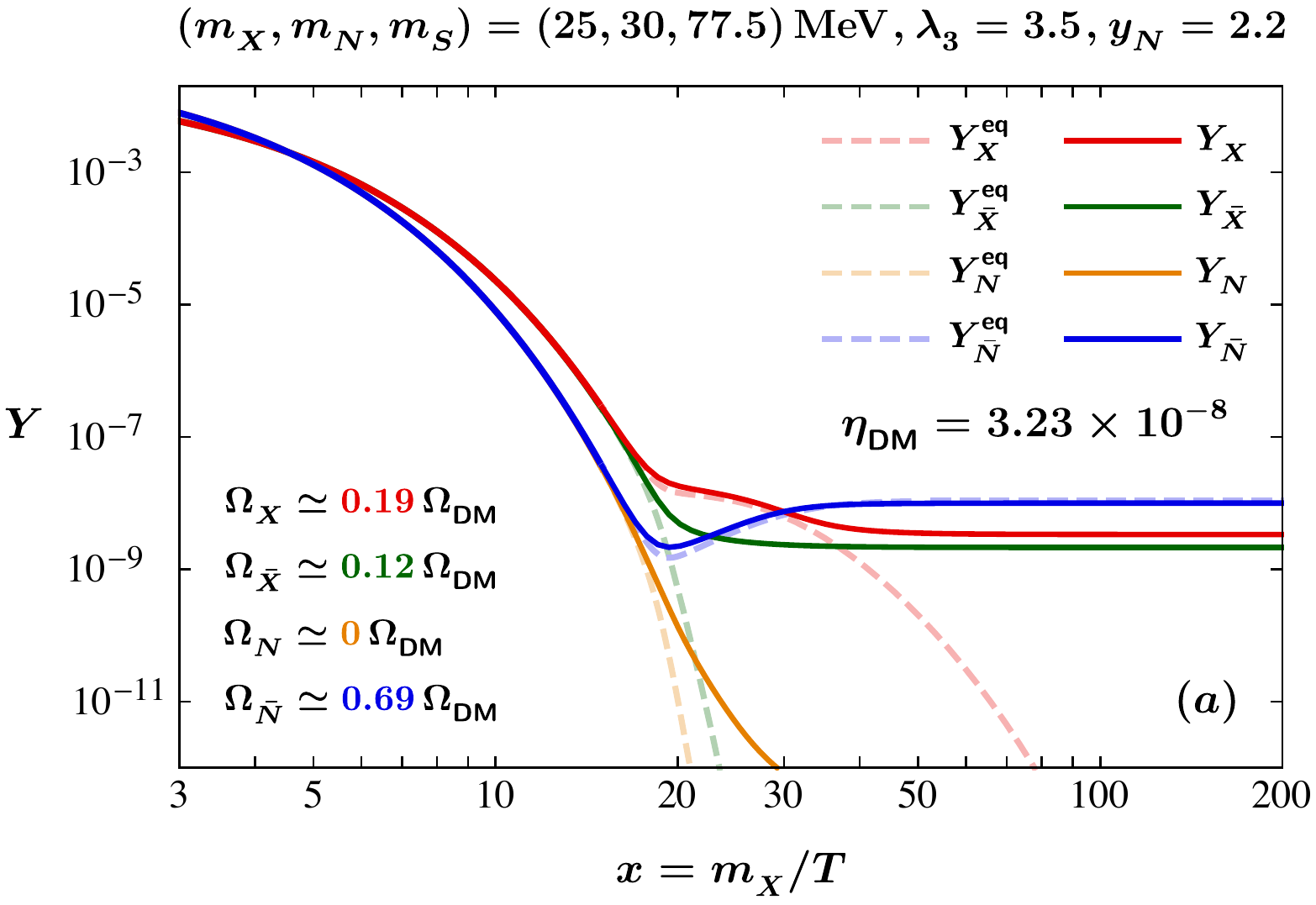}
\hs{0.0cm}
\includegraphics[width=0.49\textwidth]{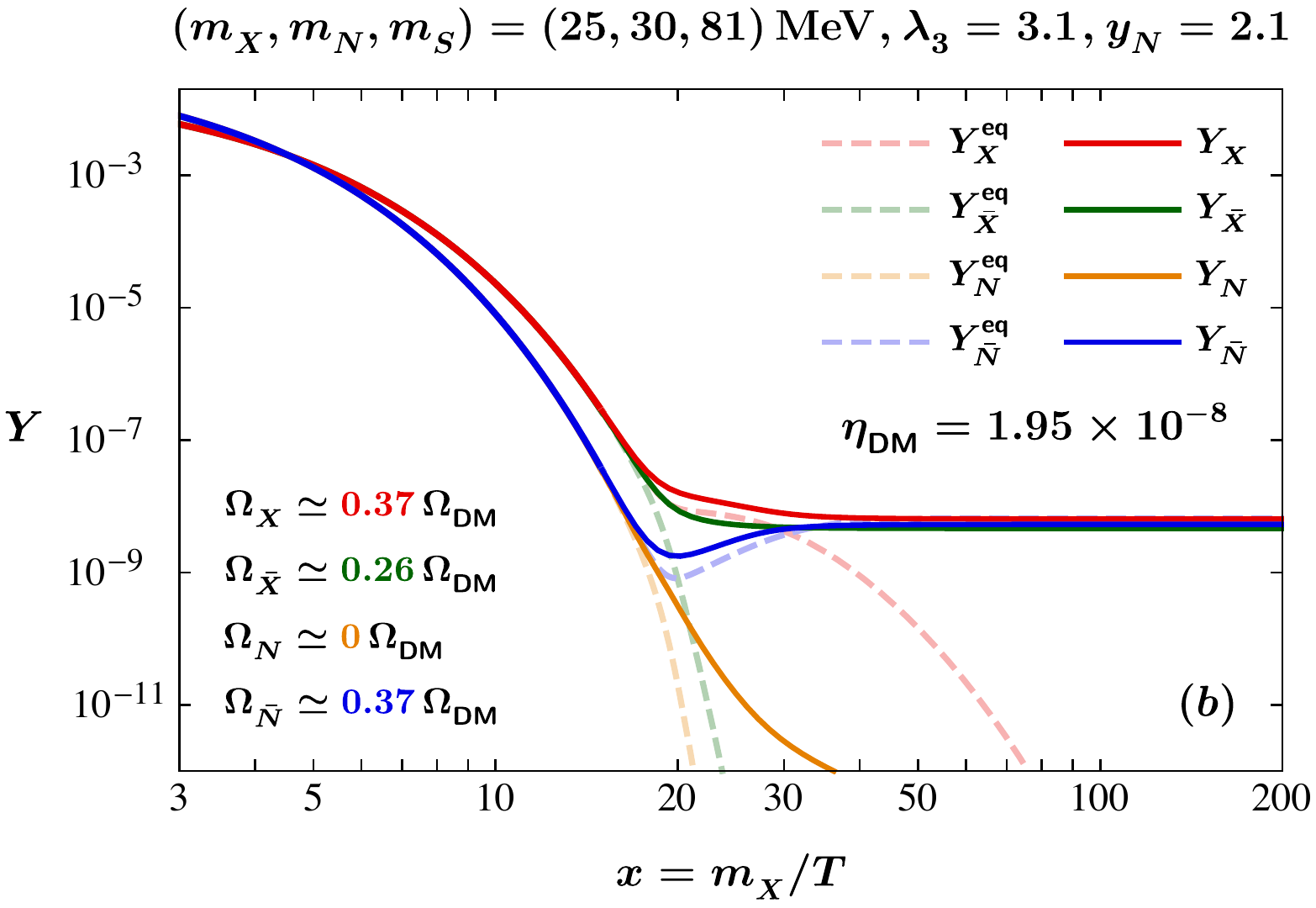}
\\[0.7cm]
\includegraphics[width=0.49\textwidth]{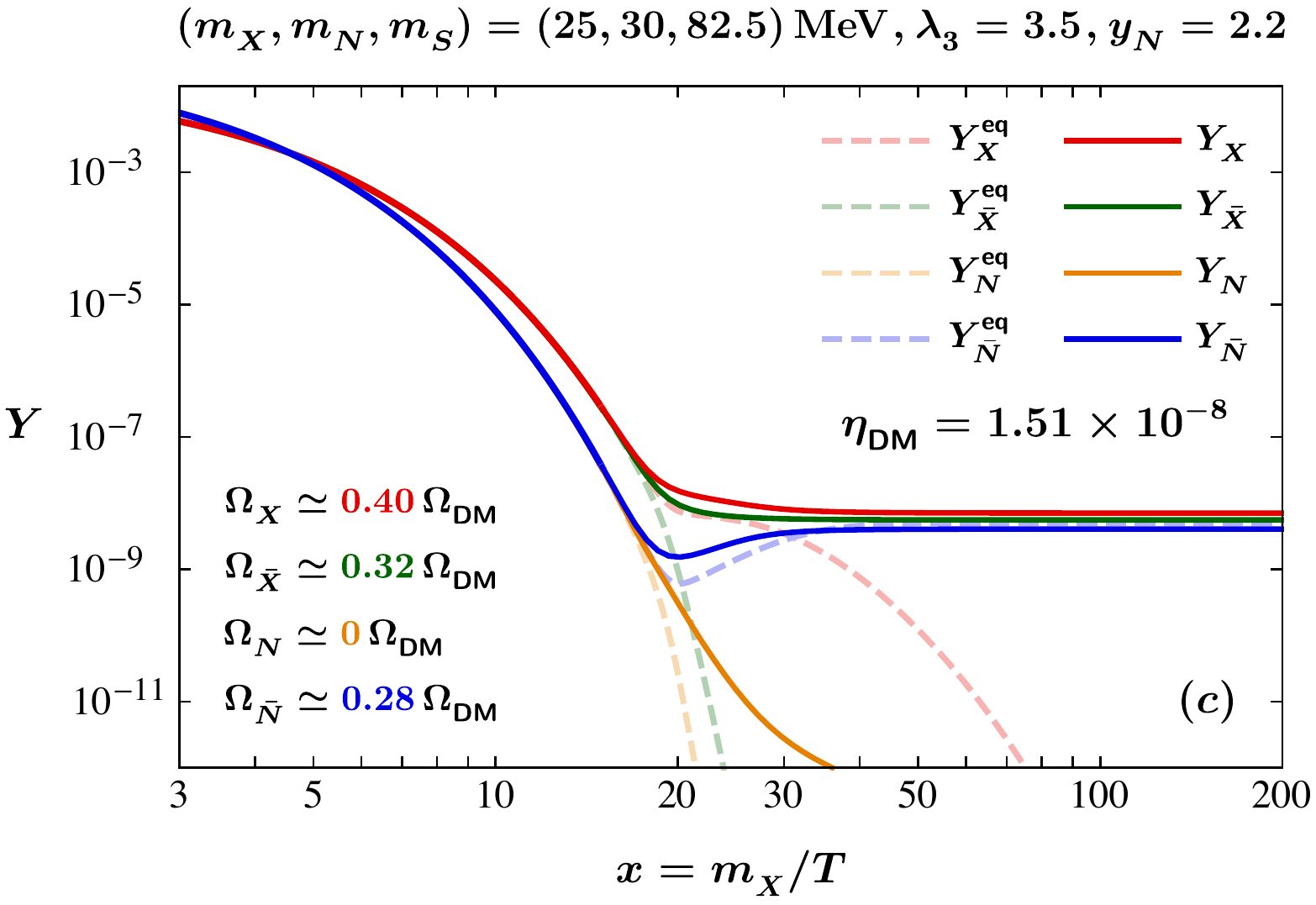}
\hs{0.0cm}
\includegraphics[width=0.49\textwidth]{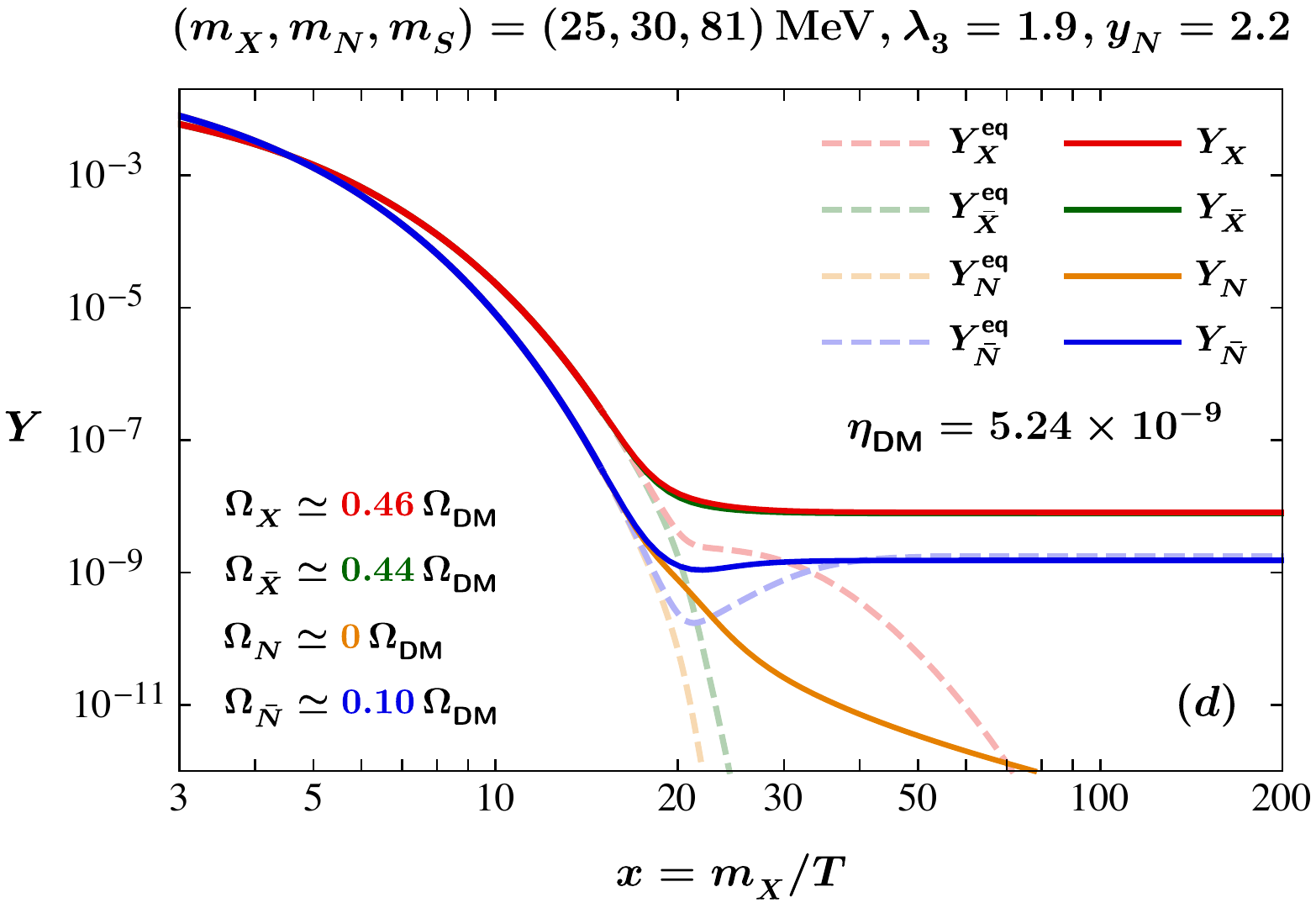}
\vs{-0.3cm}
\caption{Cosmological evolution of the comoving number densities of DM for $r^{}_N > 1$ in the $a$SIMP model, where the color solid (dashed) curves are actual (equilibrium) number densities of DM, and we have fixed the $\epsilon = 10^{-3}, m^{}_{Z'} = 250\,\tx{MeV}$, and $\lambdaXS^{} = 2.5$ in each plot.}
\label{fig:YNX}
\end{figure}

We show in Fig.\,\ref{fig:YNX} some typical time evolutions of the comoving number yields of DM in the case of $m^{}_N > m^{}_X$, where the color solid curves satisfy the observed DM relic abundance.\,\,As is evident, there are three cases of the present comoving number densities of DM, (i) $\YNb > \YX > \YXb > \YN$ (ii) $ \YX > \YNb > \YXb > \YN$ (iii) $ \YX > \YXb > \YNb > \YN$, depending on the couplings and masses of DM, and dark asymmetry.\,\,It is known that the annihilation of a fermionic DM pair via the vector portal is $s{}^{}{}^{}$-wave dominant and in our model $\langle \sigma v \rangle_{\hs{-0.03cm}\NN \to e^+e^-}\simeq 10^{-27}\,\tx{cm}^3\,\tx{s}^{-1}$ with $g^{}_\tf{D} \epsilon \simeq 10^{-3}$ and $m^{}_{Z'} = 250\,\tx{MeV}$.\,\,It follows from Eq.\,\eqref{mDMCMB} that the DM mass must be larger than a few hundred MeV.\footnote{Note that since the $\langle \sigma v \rangle_{\hs{-0.03cm}\NN \to e^+e^-} \hs{-0.05cm} \propto g_\tf{D}^2 \epsilon^2$, according to the SIMP condition in Eq.\,\eqref{gDlower} and the perturbative bound of $g^{}_\tf{D}\hs{-0.08cm}$ in Eq.\,\eqref{pertur}, we cannot make $\epsilon$ very small to evade the CMB constraint.\,\,On the other hand, the constraint of $\epsilon$ from the Belle II experiment is insensitive to $g^{}_\tf{D}\hs{-0.08cm}$ if $Z'$ mainly decays into the invisible particles.}\,\,However, as we can see in these plots, the vector-like fermions $N$ and $\bar{N}$ are highly asymmetric, $\etaN \simeq \YNb^{} \gg \YN^{}{}^{}$, as in the strong regime of the ADM scenario~\cite{Graesser:2011wi}.\,\,Thus, the severe CMB constraint on the DM mass can be alleviated.\footnote{In Ref.\,\cite{Ho:2022erb}, the vector-like fermions are symmetric.\,\,Although there is no detailed analysis of the CMB constraint for multi-component DM scenarios.\,\,However, the benchmark points for $r^{}_N <1$ in Ref.\,\cite{Ho:2022erb} may still be subject to the CMB constraint.}\,\,On the other hand, the scalar DM pair annihilation cross section through the vector portal is $p{}^{}{}^{}$-wave which is suppressed at low temperatures.\,\,Hence, the constraint from the CMB can be evaded even if the complex scalars $X$ and $\bar{X}$ belong to the intermediate regime of the ADM scenario~\cite{Graesser:2011wi}, where $\etaX \simeq \YX^{} \simeq \YXb^{}$.

In these figures, one can also observe that the number density of the vector-like fermion $\bar{N}$ increases right after the chemical freeze-out of DM.\,\,This behavior of the DM number density is called the bouncing effect of DM and has been discussed in Ref.\,\cite{Ho:2022erb}.\footnote{A more detailed discussion of the bouncing effect of DM can be found in recent paper \cite{Puetter:2022ucx}.}\,\,Here we succinctly explain this phenomenon in the following.\,\,At high temperatures (${}^{}{}^{}T>m^{}_\tf{DM}$), the DM number changing processes such as $X\hs{-0.03cm}X\hs{-0.03cm}X \hs{-0.05cm} \leftrightarrow \hs{-0.05cm} \bar{N}\hs{-0.03cm}\bar{N}$ and $X\hs{-0.03cm}X\hs{-0.03cm}N \hs{-0.05cm} \leftrightarrow \hs{-0.05cm}\bar{X}\hs{-0.03cm}\bar{N}$ maintain the chemical equilibrium of DM.\,\,As a result, the actual DM number densities track the equilibrium DM number densities.\,\,Around the freeze-out temperature of DM (${}^{}{}^{}T \simeq m^{}_\tf{DM}/20$), the backward $2 \to 3$ process, $\bar{N}\hs{-0.03cm}\bar{N} \hs{-0.05cm} \to \hs{-0.05cm} X\hs{-0.03cm}X\hs{-0.03cm}X$ becomes inactive due to the Boltzmann suppression.\,\,Then, the 
forward $3 \to 2$ process $X\hs{-0.03cm}X\hs{-0.03cm}X \hs{-0.05cm} \to \hs{-0.05cm} \bar{N}\hs{-0.03cm}\bar{N}$ starts to produce (annihilate) the vector-like fermion $\bar{N}$ (complex scalar $X$), the number of $\bar{N}$ is increased (decreased) and freezes in at low temperatures (${}^{}{}^{}T < m^{}_\tf{DM}$).\,\,Notice that since we consider the resonant mass region, where $r^{}_S \simeq 3$, then $\XXXNbNb$ is the dominant $3 \to 2$ annihilation process.\,\,Therefore, the strength of the bouncing effect increases as the level of the resonant effect increases and can be seen by comparing Fig.\,\ref{fig:YNX}${}^{}$(a) and Fig.\,\ref{fig:YNX}${}^{}$(c).\footnote{The conjugate $3 \to 2$ process $\XbXbXbNN$ with the same reaction rate as $\XXXNbNb$ can also produce the $N$ after the chemical freeze-out of DM.\,\,However, the behavior of the increasing number density of $N$ is not evident as it decreases fastly after the freeze-out temperature due to the asymmetry between $N$ and $\bar{N}$.}\,\,This bouncing effect has another advantage for this model.\,\,Typically, a complex scalar SIMP DM may have a sizable self-interacting cross section which is inconsistent with the astrophysical observations from the Bullet and Abell 3827 clusters~\cite{Markevitch:2003at,Clowe:2003tk,Massey:2015dkw,Kahlhoefer:2015vua}.\,\,However, as shown in Fig.\,\ref{fig:YNX}${}^{}$(a), the abundance of the complex scalar DM can be subdominant to that of the total DM thanks to the bouncing mechanism.\,\,In this case, the astrophysical constraints on the $a$SIMP model can be relaxed.\,\,We will discuss more details in the next section.\,\,Finally, since in Fig.\,\ref{fig:YNX}${}^{}$(a) the vector-like fermions are the dominant DM component and are extremely asymmetric, one can estimate the ratio of the DM energy density to the baryonic matter energy density for this case, where
\begin{eqnarray}
\frac{\Omega^{}_\tf{DM}}{\Omega^{}_\tf{B}}\bigg|^{}_{\tx{Fig}.\,5(a)}
\,\simeq\,
\frac{m^{}_N {}^{} \eta^{}_N + m^{}_X {}^{} \eta^{}_X}{m^{}_p {}^{} \eta^{}_\tf{B}}
\,\simeq\, 4
\end{eqnarray}
with $m_p \simeq 0.938\,\tx{GeV}$, and $\etaB \simeq 8.8 \times 10^{-11}$ the baryon number asymmetry~\cite{Graesser:2011wi}.\,\,Therefore, in the $a$SIMP scenario with a much stronger bouncing effect of DM, we can say that the value of $\Omega^{}_\tf{DM}/\Omega^{}_\tf{B}$ originates from the matter asymmetries produced in the early universe.

\begin{figure}[t!]
\centering
\includegraphics[width=0.49\textwidth]{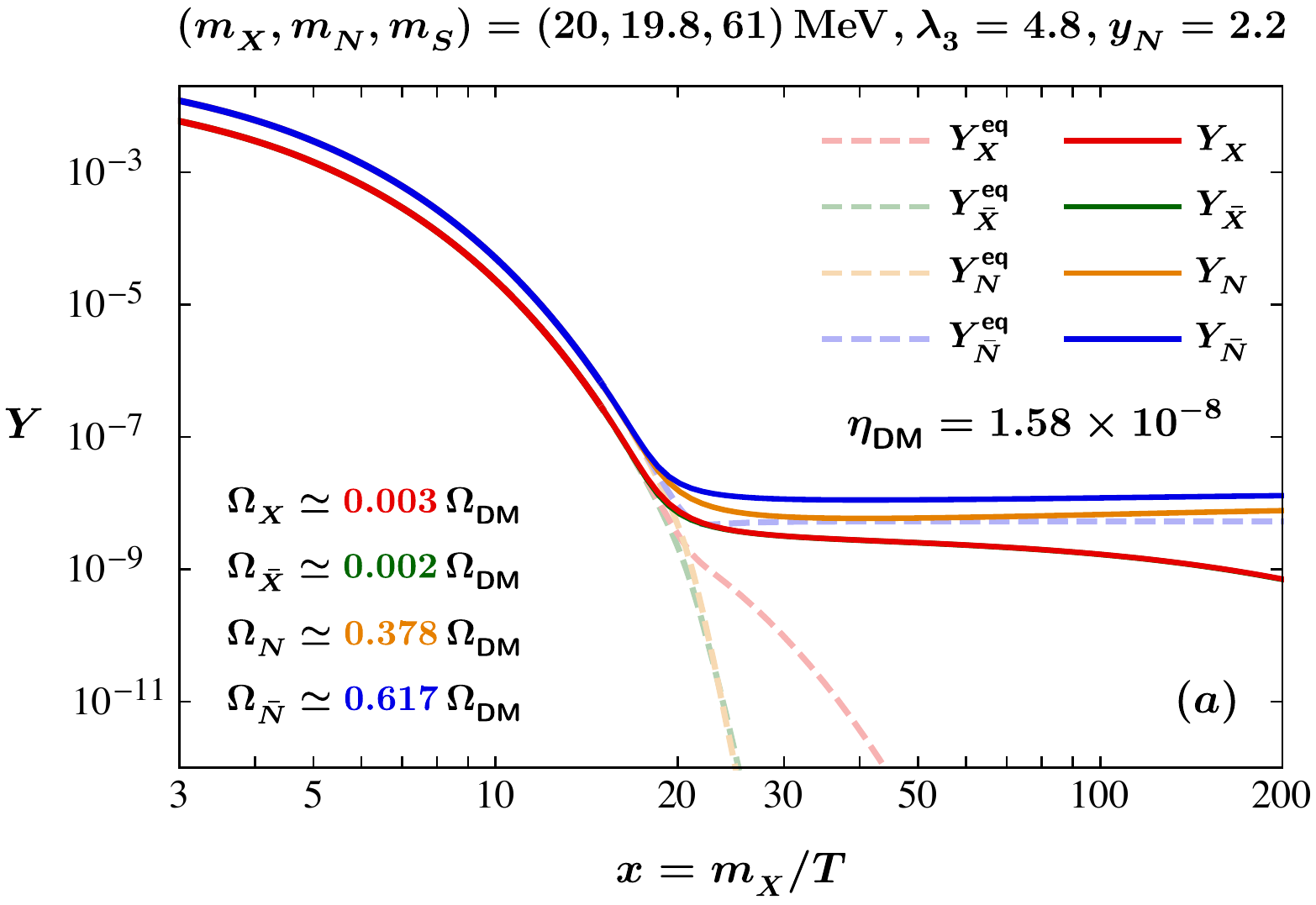}
\includegraphics[width=0.49\textwidth]{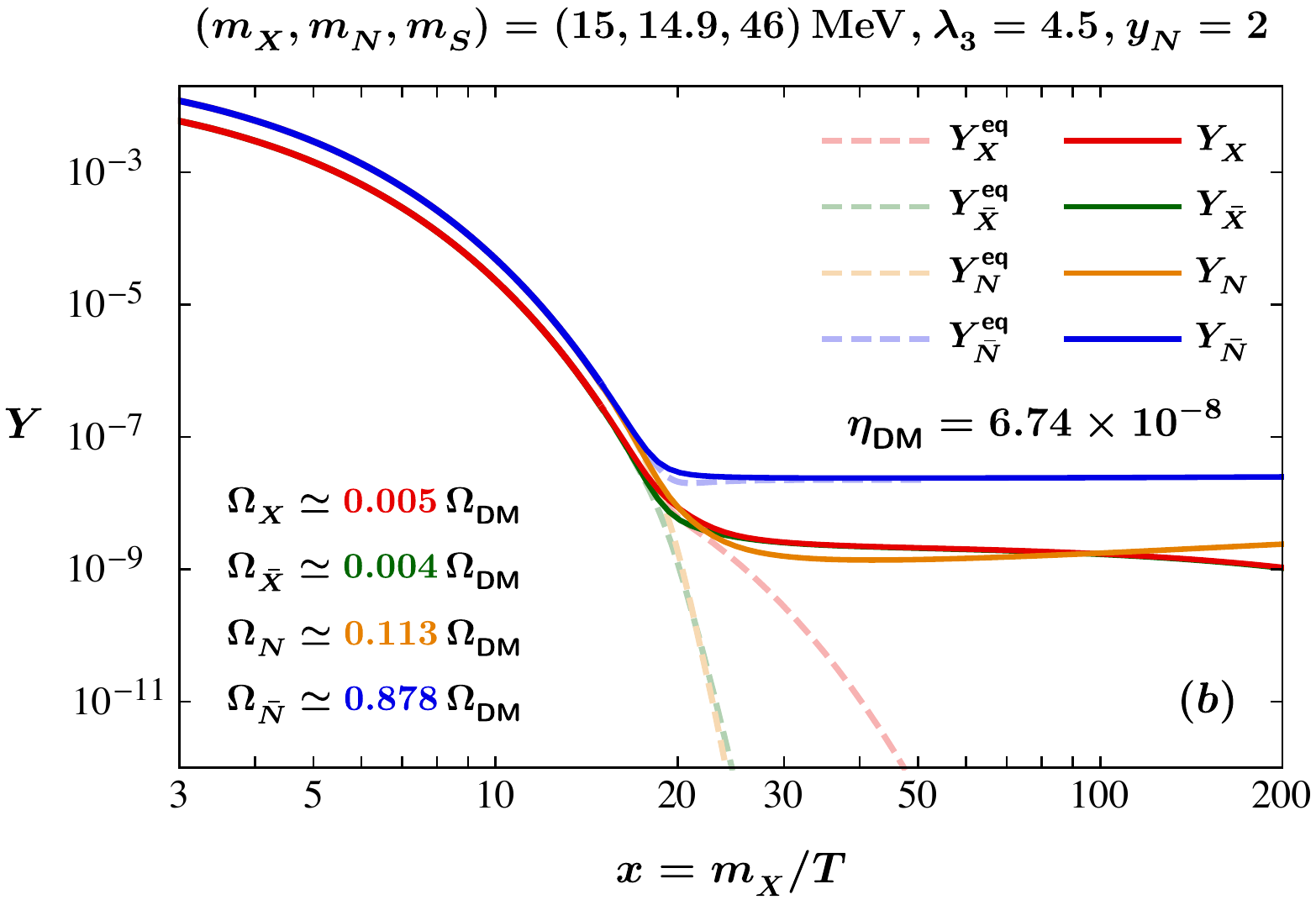}
\vs{-0.3cm}
\caption{Time evolution of the comoving number yields of DM for $r^{}_N < 1$ in the $a$SIMP model, where we have chosen the same inputs of $\epsilon{}^{}, m^{}_{Z'}$, and $\lambdaXS^{}$ as in Fig.\,\ref{fig:YNX}.}
\label{fig:YXN}
\end{figure}

Next, we show in Fig.\,\ref{fig:YXN} two examples of the time evolutions of the comoving number yields of DM in the case of $m^{}_X > m_N^{}$.\,\,As pointed out, there is only one case of the present comoving number densities of DM, $\YNb > \YN > \YX \gtrsim \YXb$.\,\,This is simple to understand since $m^{}_X > m_N^{}$ and $g^{}_N = 2{}^{}g^{}_X$, the number densities of the vector-like fermions are always bigger than that of the complex scalars during the chemical equilibrium.\,\,Plus, the number densities of the complex scalars drop further after the DM freeze-out temperature because of the $3 \to 2$ annihilations, $\XXXNbNb$ and $\XbXbXbNN$.\,\,In these figures, we can see that the vector-like fermions are not that asymmetric in comparison with Fig.\,\ref{fig:YNX}.\,\,Hence, the DM masses may be subject to the CMB constraint even in Fig.\,\ref{fig:YXN}${}^{}$(b).\,\,On the other hand, since the complex scalars have almost no relic abundances in Fig.\,\ref{fig:YXN}, the prediction of the DM self-interacting cross section would be too small to be compatible with the astrophysical observations.\,\,Also, our numerical results indicate that we have to choose degenerate DM masses to satisfy the observed DM relic density.\,\,Based on these reasons, the $m^{}_X > m^{}_N$ case is disfavored, and we will take Fig.\,\ref{fig:YNX}${}^{}$(a) as our benchmark example in the $a$SIMP model as it can satisfy all the constraints and observations, and explain the value of $\,\Omega^{}_\tf{DM}/\Omega^{}_\tf{B}$ by the asymmetries.\,\,See the next section for benchmark points.

Now, we comment on the effect of the $2 \to 2$ processes in this model.\,\,Firstly, although the $2 \to 2$ annihilations conserve the total DM number, however, each number of DM components  changes via the $2 \to 2$ processes.\,Thus, the $2 \to 2$ annihilations can also reinforce the chemical equilibrium of DM around the freeze-out temperature.\,\,In the absence of the $2 \to 2$ processes, the relic abundance of DM is overproduced for given couplings and masses of DM.\,\,Moreover, the $2 \to 2$ process $\NNbXXb$ also causes the vector-like fermions to be fully asymmetric since the $N$ can still find significant $\bar{N}$ to annihilate after the freeze-out temperature of DM.\,\,Next, let us discuss the contributions of the Feynman diagrams to the $2 \to 2$ processes.\,\,For the tree-level diagrams, with the minimum values of $g^{}_\tf{D}$, we have checked that they only affect the predicted DM density by less than 1\%, which agrees with the discussion mentioned in Ref.\,\cite{Ho:2022erb}.\,\,For the one-loop diagrams, since the $\lambdaXS$ is nothing to do with the $3 \to 2$ annihilations, we can naively turn it off for simplicity.\,\,However, a nonzero $\lambdaXS$ can strengthen the $2 \to 2$ processes to reduce the values of $\lambda^{}_3$ and $y^{}_N$, while relaxing the upper bound of $\lambda^{}_3$ from vacuum stability.\,\,Lastly, because the two-loop diagrams connect tightly to the diagrams of the $3 \to 2$ processes, they are irreducible in the $a$SIMP model.\,\,The importance of such inevitable two-loop induced diagrams for the $2 \to 2$ processes has been emphasized in Ref.\,\cite{Ho:2022erb}.

\section{Self-interacting cross section \& direct detection of DM}\label{sec:6}

In this model, both complex scalar and vector-like fermion can have self-interactions via the contact diagrams and the $S$-mediated diagrams, respectively.\footnote{By naive dimensional analysis, the DM self-interacting cross section via the $Z'$-mediated diagrams is $\sigma^{Z'}_\tf{DM} \sim (g^{}_\tf{D}/m^{}_\tf{DM})^2(m^{}_\tf{DM}/m^{}_{Z'})^4$ which is suppressed due to the small $g^{}_\tf{D}\hs{-0.08cm}$ value and heavy $Z'$ mass in this model.}\,\,Since the $N$ has almost no abundance in our benchmark example and the self-interaction of $\bar{N}$ is $d{}^{}$-wave suppressed, the complex scalar DM would mainly contribute to the DM self-interaction.\,\,Therefore, we fairly define the DM self-interacting cross section per DM mass for our benchmark example as 
\begin{eqnarray}
\frac{\sigma^{}_{\hs{-0.03cm}\tf{DM}}}{m^{}_\tf{DM}} 
\Eq
\frac{1}{m^{}_X}
\Big({}^{}
{\cal R}_{\white{\bar{\black{X}}}}^2 {}^{} 
\sigma^{}_{\hs{-0.03cm}{\white{\bar{\black{X}}}}\hs{-0.03cm}X \to {\white{\bar{\black{X}}}}\hs{-0.03cm}X} +
{\cal R}^{}_{\white{\bar{\black{X}}}} {\cal R}^{}_{\bar{X}} {}^{} 
\sigma^{}_{\hs{-0.03cm}X\hs{-0.03cm}\bar{X} \to X\hs{-0.03cm}\bar{X}} +
{\cal R}_{\bar{X}}^2 {}^{} 
\sigma^{}_{\hs{-0.03cm}\bar{X}\hs{-0.03cm}\bar{X} \to \bar{X}\hs{-0.03cm}\bar{X}} 
\Big)
~,
\end{eqnarray}
where ${\cal R}_j \,=\, \Omega_j/\Omega^{}_\tf{DM}$ is the DM fraction, and the self-interacting cross sections of the complex scalar DM are given by
\begin{eqnarray}
\sigma^{}_{\hs{-0.03cm}{\white{\bar{\black{X}}}}\hs{-0.03cm}X \to {\white{\bar{\black{X}}}}\hs{-0.03cm}X}
\,=\,
\sigma^{}_{\hs{-0.03cm}\bar{X}\hs{-0.03cm}\bar{X} \to \bar{X}\hs{-0.03cm}\bar{X}} 
\,=\,
\frac{\lambda_X^2}{8{}^{}\pi{}^{}m_X^2}
~,\quad
\sigma^{}_{\hs{-0.03cm}X\hs{-0.03cm}\bar{X} \to X\hs{-0.03cm}\bar{X}}
\,=\,
\frac{\lambda_X^2}{4{}^{}\pi{}^{}m_X^2}
~.
\end{eqnarray}

To ease some tensions between collisionless DM N-body simulations and the astrophysical observations at small-scale structures of the universe, several analyses have placed bounds on the self-interacting cross section of DM.\,\,At Milky Way and cluster scales, the self-interacting cross section of DM is within $0.1\,\tx{cm}^2/\tx{g}  \lesssim \sigma^{}_\tf{DM}/m^{}_\tf{DM} \lesssim 1\,\tx{cm}^2/\tx{g}$~\cite{Tulin:2017ara}, and the Bullet cluster provides a similar upper bound, where 
$\sigma^{}_\tf{DM}/m^{}_\tf{DM} \lesssim 1\,\tx{cm}^2/\tx{g}$~\cite{Markevitch:2003at,Clowe:2003tk}.\,\,At the same time, the Abell 3827 cluster gives $1\,\tx{cm}^2/\tx{g} \lesssim \sigma^{}_\tf{DM}/m^{}_\tf{DM} \lesssim 3\,\tx{cm}^2/\tx{g}$~\cite{Massey:2015dkw,Kahlhoefer:2015vua} which does not overlap with the aforementioned two restriction ranges.\,\,More recent observations on cluster collisions have led to the strongest upper bound on $\sigma^{}_\tf{DM}/m^{}_\tf{DM} \lesssim 0.47\,\tx{cm}^2/\tx{g}$~\cite{Harvey:2015hha}.

We show our predictions of the DM self-interacting cross section for a few benchmark points with $m^{}_N \hs{-0.04cm} > \hs{-0.02cm} m^{}_X$ in Tab.\,\ref{tab:2}, where we have considered an optimistic value for the kinetic mixing parameter, $\epsilon = 10^{-3}$.\footnote{One can take the slightly smaller $\epsilon$ value such that the $g^{}_\tf{D}\hs{-0.08cm}$ value can be slightly larger according to Eq.\,\eqref{gDlower}.\,\,As a consequence, the tree-level $2 \to 2$ process $\NNbXXb$ via the $Z'$-mediated diagram may affect the predicted DM relic density by few percent.\,\,However, one cannot choose too small $\epsilon$, otherwise, the perturbative bound of $g^{}_\tf{D}\hs{-0.08cm}$ would be violated based on our charge assignment in Tab.\,\ref{tab:1}.}\,\,As expected, we see that the values of $\sigma^{}_\tf{DM}/m^{}_\tf{DM}$ can be compatible with the above-mentioned astrophysical bounds, especially the third benchmark point.

\begin{table}[t!]
\begin{center}
\def\arraystretch{1.2}
\begin{tabular}{|c|c|c|c|c|c|c|c|c|}
\hline
~$\lambda^{}_X$~ & ~$\lambda^{}_S$~ & ~$\lambda^{}_{X\hs{-0.03cm}S}$~ &~$\lambda^{}_3$~ & ~$y^{}_N$~ & ~$\etaDM$~ & ~$\big(m^{}_X,m^{}_N,m^{}_S\big)/\tx{MeV}$~ & ~$\sigma^{}_\tf{DM}/m^{}_\tf{DM}\,(\tx{cm}^2/\tx{g})$~ & ~$\sigma^{}_e\,(\tx{cm}^2)$~ 
\\[0.05cm]
\hline 
~$4.2$~ & ~$4.5$~ & ~$0.0$~ & ~$3.6$~ & ~$2.2$~ & ~$3.23 \times 10^{-8}$~ &~$(25,30,77.5)$~ & ~$0.96$~ & ~$7.47 \times 10^{-42}$~ 
\\\hline  
~$3.5$~ & ~$2.5$~ & ~$1.0$~ & ~$3.1$~ & ~$2.1$~ & ~$4.05 \times 10^{-8}$~ &~$(20,25,61)$~ & ~$1.06$~ & ~$1.52 \times 10^{-41}$~ 
\\\hline  
~$2.9$~ & ~$2.3$~ & ~$2.5$~ & ~$3.2$~ & ~$1.5$~ & ~$4.46 \times 10^{-8}$~ &~$(20,24,61)$~ & ~$0.43$~ & ~$1.54 \times 10^{-41}$~ 
\\\hline  
~$3.9$~ & ~$2.1$~ & ~$3.5$~ & ~$4.2$~ & ~$2.5$~ & ~$4.06 \times 10^{-8}$~ &~$(20,24,59.5)$~ & ~$1.53$~ & ~$1.46 \times 10^{-41}$~ 
\\\hline  
\end{tabular}
\caption{Benchmark points for $r^{}_N > 1$ in the $a$SIMP model.}
\label{tab:2}
\end{center}
\vs{-0.5cm}
\end{table}

As shown in Fig.\,\ref{fig:YNX} and Tab.\,\ref{tab:2}, the preferred DM mass scale in the $a$SIMP model is around ${\cal O}(20)$\,MeV.\,\,For these DM masses, one can use DM scattering off an electron inside the atom to detect the DM particles in this model.\,\,Utilizing the gauge interactions in Eq.\,\eqref{Zprime}, the total DM-$e^-$ elastic scattering cross section is computed as
\begin{eqnarray}
\sigma^{}_e
\,=\,
\big({}^{}{\cal R}^{}_{\white{\bar{\black{N}}}} + {\cal R}^{}_{\bar{N}} \big) {}^{}
\sigma^{}_{\hs{-0.03cm}N\hs{-0.02cm}e \to N\hs{-0.02cm}e}
+
\big({}^{}{\cal R}^{}_{\white{\bar{\black{X}}}} + {\cal R}^{}_{\bar{X}} \big) {}^{}
\sigma^{}_{\hs{-0.03cm}X\hs{-0.02cm}e \to X\hs{-0.02cm}e}
~,
\end{eqnarray}
where the individual DM-$e^{-}$ scattering cross sections are given by
\begin{eqnarray}
\sigma^{}_{\hs{-0.03cm}N\hs{-0.02cm}e \to N\hs{-0.02cm}e}
\,=\,
\frac{c_{N\hs{-0.02cm}e}^2}{\pi}\frac{\mu_{N\hs{-0.02cm}e}^2}{m_{Z'}^4}
~,\quad
\sigma^{}_{\hs{-0.03cm}X\hs{-0.02cm}e \to X\hs{-0.02cm}e}
\,=\,
\frac{c_{X\hs{-0.02cm}e}^2}{\pi}\frac{\mu_{X\hs{-0.02cm}e}^2}{m_{Z'}^4}
\end{eqnarray}
with $c^{}_{je}$ the product of the gauge couplings, and $\mu^{}_{je}$ the reduced mass of the DM-$e^{-}$ system
\begin{eqnarray}
c^{}_{je} \,=\, g^{}_\tf{D}{}^{}g^{}_\tf{e}{}^{}c^{}_\tf{W}{}^{}\epsilon{}^{}{\cal Q}^{}_j
~,\quad
\mu^{}_{je} \,=\, \frac{m^{}_j m^{}_e}{m^{}_j + m^{}_e}
~,
\end{eqnarray}
where $m^{}_e$ is the electron mass.

We show our predictions of the DM-$e^-$ elastic scattering cross section in the last column of Tab.\,\ref{tab:2} for each benchmark point, where we have fixed $g^{}_\tf{D} \epsilon$ to the maginal values in Eq.\,\eqref{gDlower} with $m^{}_{Z'} = 250\,\tx{MeV}$.\,\,The current available upper bounds for the DM-$e^-$ elastic scattering cross section come from XENON1T~\cite{XENON:2019gfn}, XENON10~\cite{Essig:2017kqs}, and DarkSide-50~\cite{DarkSide:2018ppu} collaborations.\,\,These experiments provide upper limits for a heavy mediator, where $\sigma^{}_e \sim 10^{-37}\,\tx{cm}^2$ to $\sim 10^{-36}\,\tx{cm}^2$ with the DM mass from $\sim 20\,\tx{MeV}$ to $\sim 30\,\tx{MeV}$.\,\,Hence, our predictions for the DM-$e^-$ elastic scattering cross section are still far below the last up-to-date sensitivities.\,\,Nevertheless, some projected experiments try to apply semiconductors~\cite{Griffin:2020lgd}, superconductors~\cite{Hochberg:2021pkt}, superconducting nanowires~\cite{Hochberg:2019cyy}, etc.\,to probe low mass DM.\,\,For the DM mass in tens of MeV, their sensitivities of the DM-$e^-$ scattering cross section can potentially reach $\sigma_e \sim 10^{-41}\,\tx{cm}^2$, which can be used to test the benchmark points in the $a$SIMP model.

\section{Discussion \& Conclusions}\label{sec:7}

\begin{figure}[t!]
\centering
\includegraphics[width=0.49\textwidth]{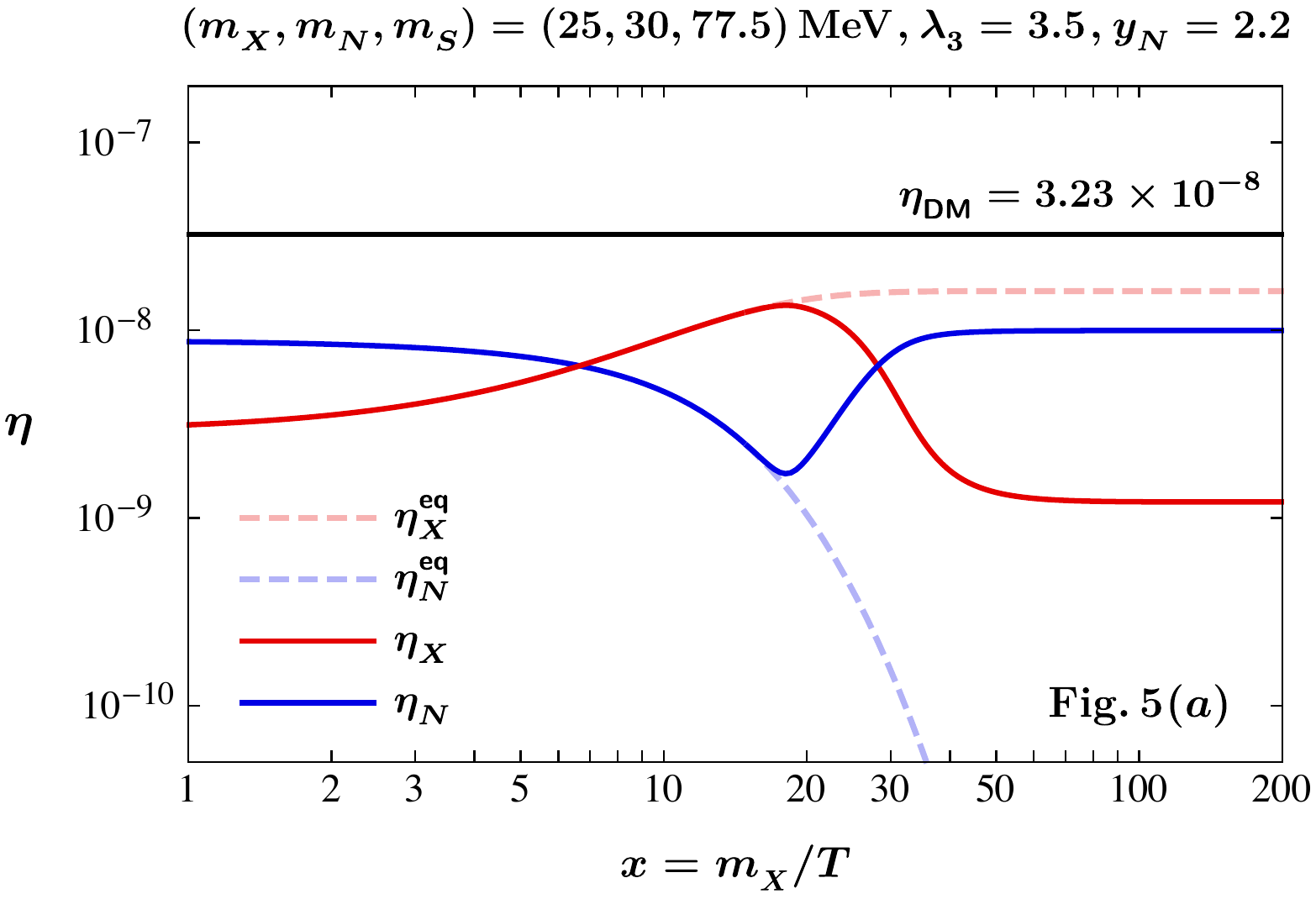}
\vs{-0.3cm}
\caption{Cosmological evolution of the dark asymmetries corresponds to benchmark figure \ref{fig:YNX}${}^{}$(a), where the color solid (dashed) curves are actual (equilibrium) DM asymmetries.\,\,Figs.\,\ref{fig:YNX}${}^{}$(b), \ref{fig:YNX}${}^{}$(c), and \ref{fig:YNX}${}^{}$(d) have a similar evolution of the dark asymmetries.}
\label{fig:etaNX}
\end{figure}

Before going to the conclusion, let us discuss the dark asymmetry in the $a$SIMP model.\,\,So far,
we have treated the $\etaDM\hs{-0.05cm}$ as a free input parameter when solving the Boltzmann equations. However, it may have a physical origin akin to the baryon asymmetry.\,\,Since the $\etaDM\hs{-0.05cm}$ is a sum of two distinct DM asymmetries, thus we have to know how these two DM asymmetries evolve with the cosmic time.\,\,We show in Fig.\,\ref{fig:etaNX} the time evolution of the DM asymmetries for Fig.\,\ref{fig:YNX}${}^{}$(a). As illustrated, the DM asymmetries are separately conserved at high and low temperatures and redistributed during the freeze-out temperature of DM.\,\,At low temperatures, the values of the DM asymmetries are determined by  numerical computations.\,\,On the other hand, the values of DM asymmetries at high temperatures can be calculated analytically.\,\,First, we can take a ratio of the equilibrium DM asymmetries.\,\,At very high temperatures, one can show that
\begin{eqnarray}
\frac{\eta^\tf{eq}_N (x)}{\eta^\tf{eq}_X (x)} \bigg|^{}_{x \to 0}
\,=\,
\frac{3}{2}
~.
\end{eqnarray}
Next, with $\etaDM = 3{}^{}\eta^\tf{eq}_N(x) + 2{}^{}\eta^\tf{eq}_X(x)$, we then obtain $\eta^\tf{eq}_N = 3{}^{}\etaDM/13$ and $\eta^\tf{eq}_X = 2{}^{}\etaDM/13$, which is consistent with our numerical result.\,\,These relations imply that the typical order of the DM asymmetries in the $a$SIMP scenario is $\eta^\tf{eq}_{N,X} \sim 10^{-9}$ to $10^{-8} \gg \etaB$.\,\,Hence, we can generate these DM asymmetries using the same process as baryogenesis or leptogenesis.\,\,For instance, one can introduce a dark number violating interaction as ${\cal L}_{\cancel{\tf{DM}}} = -{}^{}{}^{}y^{}_\psi{}^{}\overline{\psi}{}^{}N{}^{}\zeta + \tx{h.c.}$, where $\psi$ is a heavy Majorana fermion, $\zeta$ is a complex scalar, and $y^{}_\psi$ is a complex Yukawa coupling.\,\,By defining a dark number asymmetry, $\epsilon^{}_N = \big[ \Gamma(\psi \to N \zeta) - \Gamma(\psi \to \bar{N} \zeta^\ast)\big]/\Gamma_\psi$ with $\Gamma_\psi$ the decay rate of $\psi$, we can then relate the dark number asymmetry to the dark asymmetry as $\etaN \hs{-0.05cm} \sim \epsN/g^{}_\star$.\,With $\epsN \sim y_\psi^2/(8\pi) \sim 10^{-6}$ and $g^{}_\star \sim 10^2$ to $10^3$, the right order of $\,\etaN$ can be obtained in the $a$SIMP scenario.\,\,However, the construction of a UV complete model to realize this dark asymmetry is beyond the scope of this paper, and we leave the detailed study as our future work.

In this article, we have built for the first time an asymmetric SIMP DM model, where the asymmetric DM are comprised of the vector-like fermion and complex scalar both with nonzero chemical potentials.\,\,These two DM particles are stabilized by the accidental ${}^{}{}^{}\Zf$ symmetry after the breaking of the U$(1)^{}_\tf{D}$ gauge symmetry.\,\,By introducing one extra complex scalar to link the SIMP DM, this model can have the $3 \to 2$ and $2 \to 2$ processes that determine the relic abundance of DM.\,\,In particular, the $2 \to 2$ processes can reinforce the chemical equilibrium of DM around the DM freeze-out temperature in contrast with other SIMP DM models.\,\,Also, by taking the marginal values of the dark gauge coupling, we can suppress the reaction rate of the WIMP scenario.\,\,Meanwhile, the SIMP DM can keep kinetic equilibrium with the thermal plasma sufficiently until the freeze-out time of DM such that the $a$SIMP scenario is successful.

A striking feature of the $a$SIMP model is that there can be a DM bouncing effect, by which the number density of the vector-like fermion DM can increase after the chemical freeze-out of DM.\,\,Correspondingly, the number yield of the complex scalar DM becomes subdominant, thus the prediction of the DM self-interaction cross section can be compatible with the astrophysical observations.\,\,In addition, if the vector-like fermion DM is fully asymmetric, then the total DM relic density is mainly contributed by the asymmetric component of DM.\,\,In this case, we can explain the DM-to-baryon energy density ratio by primordial matter asymmetries produced in the very early universe.\,\,Finally, we have found several benchmark points which can satisfy all the theoretical and observational constraints and predicted the DM-$e^-$ elastic scattering cross section to examine this model in future prospective experiments using electron target.

\acknowledgments

SYH would like to thank Shih-Yen Tseng for his useful discussion and collaboration in the early stage of the present work.\,\,This work is supported by KIAS Individual Grants under Grant No.\,PG081201 (SYH).

\end{document}